\documentclass[%
 reprint,
superscriptaddress,
 amsmath,amssymb,longbibliography,
 aps,
pra,
]{revtex4-1}

\usepackage{graphicx}
\usepackage{dcolumn}
\usepackage{bm}
\usepackage{hyperref}

\usepackage{mhchem}
\usepackage[export]{adjustbox}
\usepackage{subcaption}
\usepackage{afterpage}
\usepackage{mathtools}

\DeclareMathOperator{\sinc}{sinc}
\DeclarePairedDelimiter\abs{\lvert}{\rvert}

\begin{document}

\title{Space-compatible cavity-enhanced single-photon generation with hexagonal boron nitride}

\author{Tobias Vogl}
\email{Tobias.Vogl@anu.edu.au}
\affiliation{Centre for Quantum Computation and Communication Technology, Department of Quantum Science, Research School of Physics and Engineering, The Australian National University, Acton ACT 2601, Australia}
\author{Ruvi Lecamwasam}
\affiliation{Centre for Quantum Computation and Communication Technology, Department of Quantum Science, Research School of Physics and Engineering, The Australian National University, Acton ACT 2601, Australia}
\author{Ben C. Buchler}
\affiliation{Centre for Quantum Computation and Communication Technology, Department of Quantum Science, Research School of Physics and Engineering, The Australian National University, Acton ACT 2601, Australia}
\author{Yuerui Lu}
\affiliation{Centre for Quantum Computation and Communication Technology, Research School of Electrical, Energy and Materials Engineering, The Australian National University, Acton ACT 2601, Australia}
\author{Ping Koy Lam}
\email{Ping.Lam@anu.edu.au}
\affiliation{Centre for Quantum Computation and Communication Technology, Department of Quantum Science, Research School of Physics and Engineering, The Australian National University, Acton ACT 2601, Australia}

\date{\today}

\begin{abstract}
Sources of pure and indistinguishable single-photons are critical for near-future optical quantum technologies. Recently, color centers hosted by two-dimensional hexagonal boron nitride (hBN) have emerged as a promising platform for high luminosity room temperature single-photon sources. Despite the brightness of the emitters, the spectrum is rather broad and the single-photon purity is not sufficient for practical quantum information processing. Here, we report integration of such a quantum emitter hosted by hBN into a tunable optical microcavity. A small mode volume of the order of $\lambda^3$ allows us to Purcell enhance the fluorescence, with the observed excited state lifetime shortening. The cavity significantly narrows the spectrum and improves the single-photon purity by suppression of off-resonant noise. We explore practical applications by evaluating the performance of our single-photon source for quantum key distribution and quantum computing. The complete device is compact and implemented on a picoclass satellite platform, enabling future low-cost satellite-based long-distance quantum networks.
\end{abstract}

\maketitle


\section{Introduction}
Near-future optical quantum information processing\cite{10.1038/nphoton.2009.229} relies on sources of pure and indistinguishable single-photons. Promising candidates include quantum dots\cite{10.1038/nnano.2017.218}, trapped ions\cite{10.1088/1367-2630/11/10/103004}, color centers in solids\cite{10.1038/nphoton.2016.186} and single-photon sources (SPSs) based on heralded spontaneous parametric downconversion\cite{10.1364/OE.24.023992}. The recent discovery of fluorescent defects in two-dimensional (2D) materials has added yet another class of quantum emitters to the solid state color centers. Stable quantum emitters have been reported in the transition metal dichalcogenides WSe$_2$\cite{10.1364/OPTICA.2.000347,nnano.2015.60,nnano.2015.67,nnano.2015.75,nnano.2015.79}, WS$_2$\cite{10.1038/ncomms12978}, MoSe$_2$\cite{10.1063/1.4945268} and MoS$_2$\cite{arXiv:1901.01042}. The optical transition energies for these emitters, however, are located in close vicinity to the electronic band gap. Thus, cryogenic cooling below 15$\,$K is required to resolve the zero phonon lines (ZPLs). For room temperature quantum emission, defects hosted by large band gap materials are ideal, as has been demonstrated in 2D hexagonal boron nitride (hBN)\cite{nnano.2015.242,10.1021/acsnano.6b03602,10.1103/PhysRevApplied.5.034005}. In this case, the energy levels introduced by the defects into the band structure are well isolated. The large band gap of 6$\,$eV\cite{10.1038/nphoton2015.77} also prevents non-radiative decay, which in turn allows for high quantum efficiencies. Unlike for solid state quantum emitters in 3D systems, the 2D crystal lattice of hBN allows for an intrinsically high extraction efficiency. More precisely, the single-photon emitters have an in-plane dipole resulting in out-of-plane emission, where the emitters are not surrounded by high refractive index materials. Hence, total internal or Frensel reflection does not affect the collection of the single-photons. Furthermore, 2D crystals can be easily attached by Van der Waals forces to components such as fibers or waveguides, making them suitable for integration with photonic networks\cite{10.1021/acsphotonics.7b00025,10.1088/1361-6463/aa7839}. The exceptionally high thermal and chemical robustness of hBN benefits the durability of the quantum emitters, achieving long-term stable operation\cite{10.1021/acsphotonics.8b00127} over a huge temperature range\cite{10.1021/acsphotonics.7b00086}. Moreover, the quantum emitters (and 2D materials in general) have a high tolerance to ionizing radiation, allowing for use in space applications\cite{arXiv:1811.10138}.\\
\indent In spite of large experimental research efforts and theoretical calculations\cite{10.1039/C7NR04270A,10.1021/acsphotonics.7b01442}, the exact nature of the defects yet has to be determined. Furthermore, the identification is hampered by the large variations of the lifetime and ZPL wavelength from defect to defect. Lifetimes ranging from 0.3 up to 20$\,$ns\cite{10.1021/acsphotonics.8b00127,10.1021/acsphotonics.7b00025} and ZPLs in the UV\cite{10.1021/acs.nanolett.6b01368} and the full visible spectrum\cite{10.1021/acsnano.6b03602,10.1021/acsphotonics.6b00736} have been reported. In addition to naturally occurring defects\cite{nnano.2015.242}, the emitters can also be created artificially using diverse methods, including chemical etching\cite{10.1021/acs.nanolett.6b03268}, plasma etching\cite{10.1021/acsphotonics.8b00127,10.1039/C7NR08222C}, ion\cite{10.1021/acsami.6b09875} and electron irradiation\cite{10.1021/acsami.6b09875,10.1021/acsami.8b07506} as well as near-deterministic stress-induced activation\cite{10.1364/OPTICA.5.001128}. Although most researchers agree that quantum emitters in hBN provide a number of unique opportunities, the performance still lags behind state-of-the-art SPSs. Moreover, the reported quality of single-photons from hBN is not sufficient for practical quantum information processing like quantum key distribution (QKD)\cite{10.1103/RevModPhys.74.145} or photonic quantum computing\cite{10.1103/RevModPhys.79.135}.\\
\indent A straightforward path for improving the performance of a spontaneous emission process is to use the Purcell effect by coupling the emitter to an optical resonator\cite{10.1038/nature01939}. The optical resonator reduces the number of modes the emitter can couple to, thereby enhancing emission into the resonant modes. This even works in the "bad-emitter" regime, when the emitter linewidth is larger than the cavity linewidth\cite{10.1103/PhysRevA.88.053812}. Work on cavity-integration of emitters in 2D materials has been reported, with quantum emitters hosted by WSe$_2$ coupled to plasmonic nanocavities\cite{10.1364/OE.26.025944,10.1038/s41565-018-0275-z} and microcavities\cite{10.1063/1.5026779}. Quantum emitters hosted by hBN have been coupled to plasmonic nanocavities\cite{doi:10.1021/acs.nanolett.7b00444}. Hexagonal boron nitride can also be used to fabricate photonic crystal cavities, however, this makes the required spectral matching between optical cavity mode and emitter difficult\cite{10.1038/s41467-018-05117-4}. Yet, the performance is still not sufficient for use in quantum information experiments.\\
\indent In this article, we report room temperature single-photon emission from multilayer hBN flakes coupled with a microcavity. The plano-concave cavity fully suppresses the phonon sideband (PSB) and other off-resonant noise, while at the same time greatly enhances directionality and  the spontaneous emission rate. The hemisphere is fabricated using focused ion beam (FIB) milling, allowing for a small radius of the accurate and precise curvature. This leads to an ultra-small mode volume on the order of $\lambda^3$. We fully characterize the SPS and assess its feasibility for QKD and quantum computing. Moreover, the single-photon source in its current configuration is fully self-contained and compact enough for integration on a pico-class satellite, making it interesting for satellite-based quantum communication\cite{10.1038/nature23655}.

\section{Design and fabrication}
\begin{figure*}[t!]
\centering
  \includegraphics[width=0.257\linewidth,keepaspectratio,valign=t]{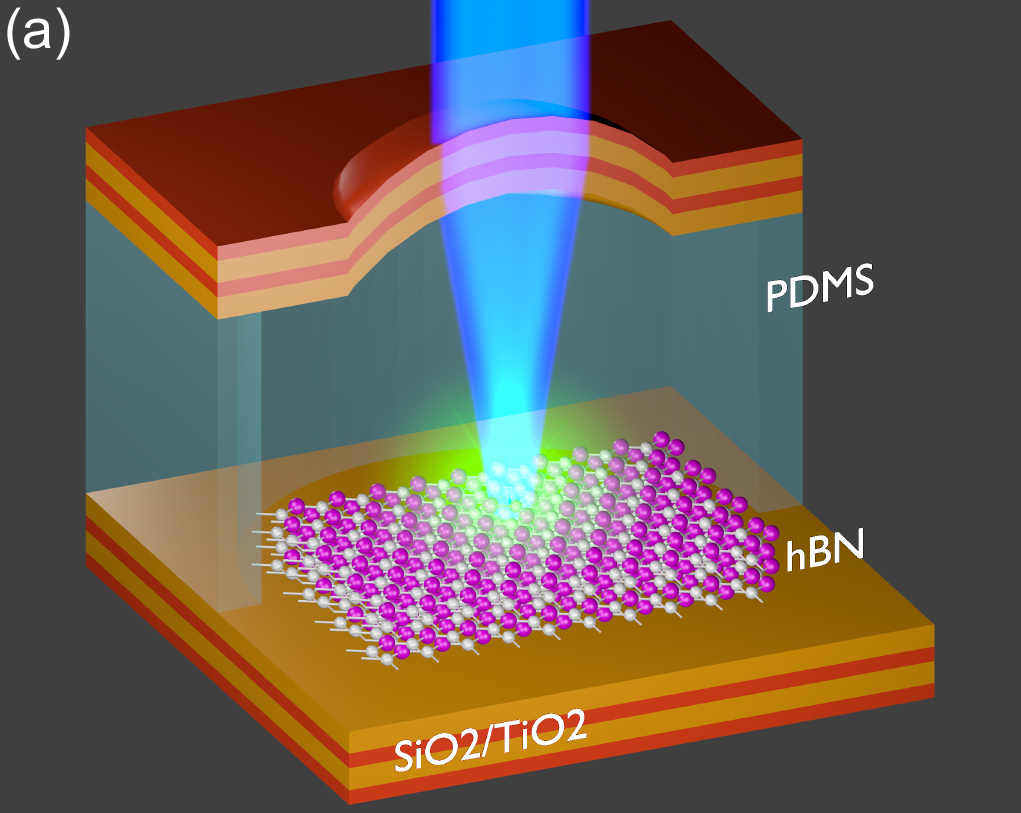}
  \includegraphics[width=0.383\linewidth,keepaspectratio,valign=t]{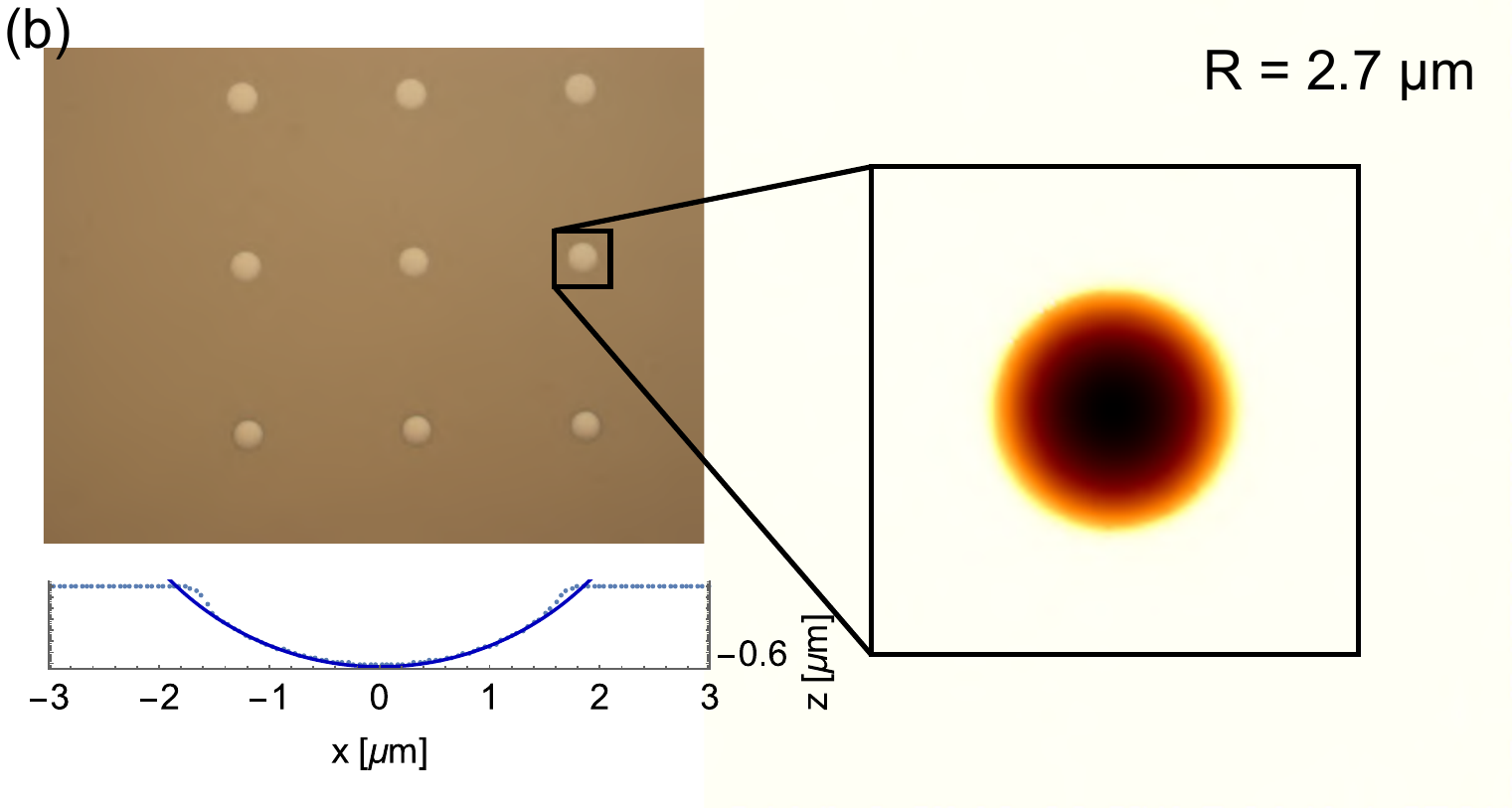}
  \includegraphics[width=0.32\linewidth,keepaspectratio,valign=t]{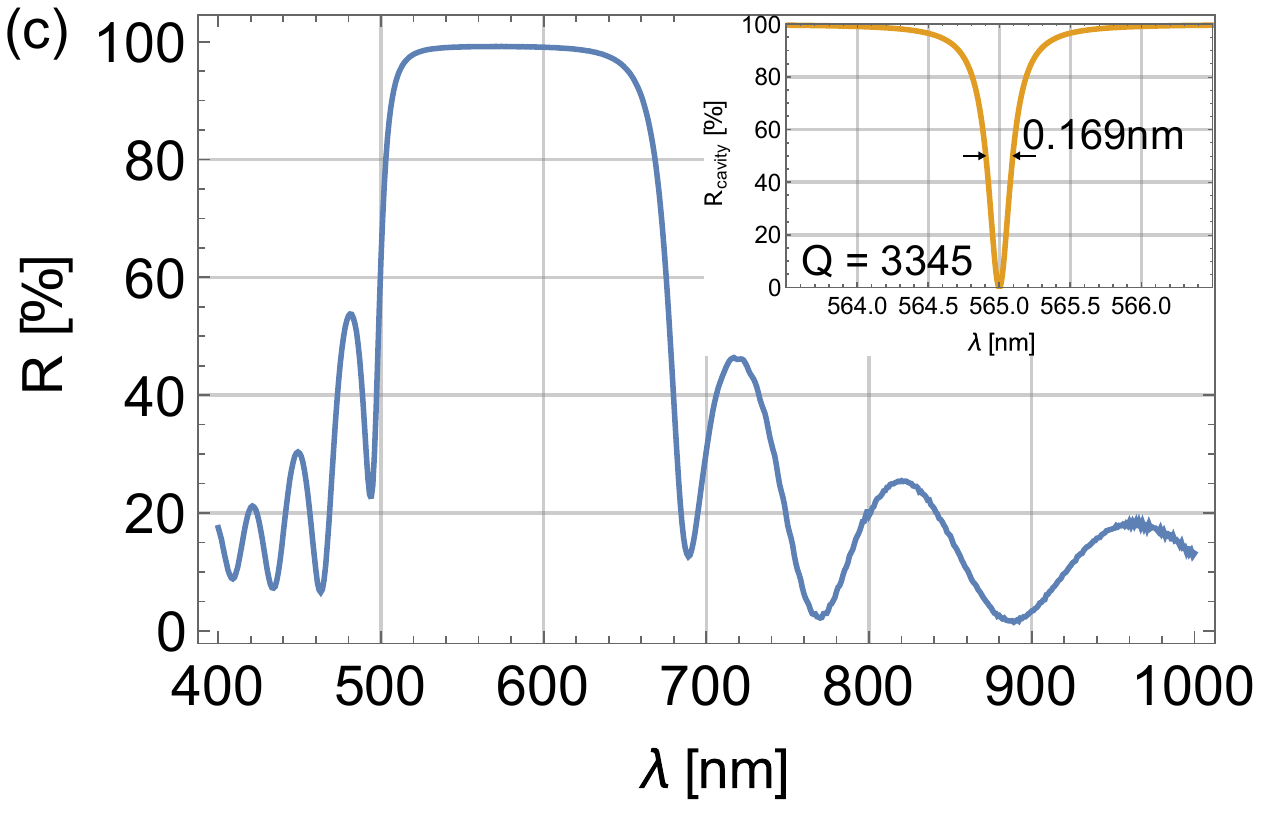}\\
  \vspace{0.1cm}
  \includegraphics[width=0.32\linewidth,keepaspectratio,valign=t]{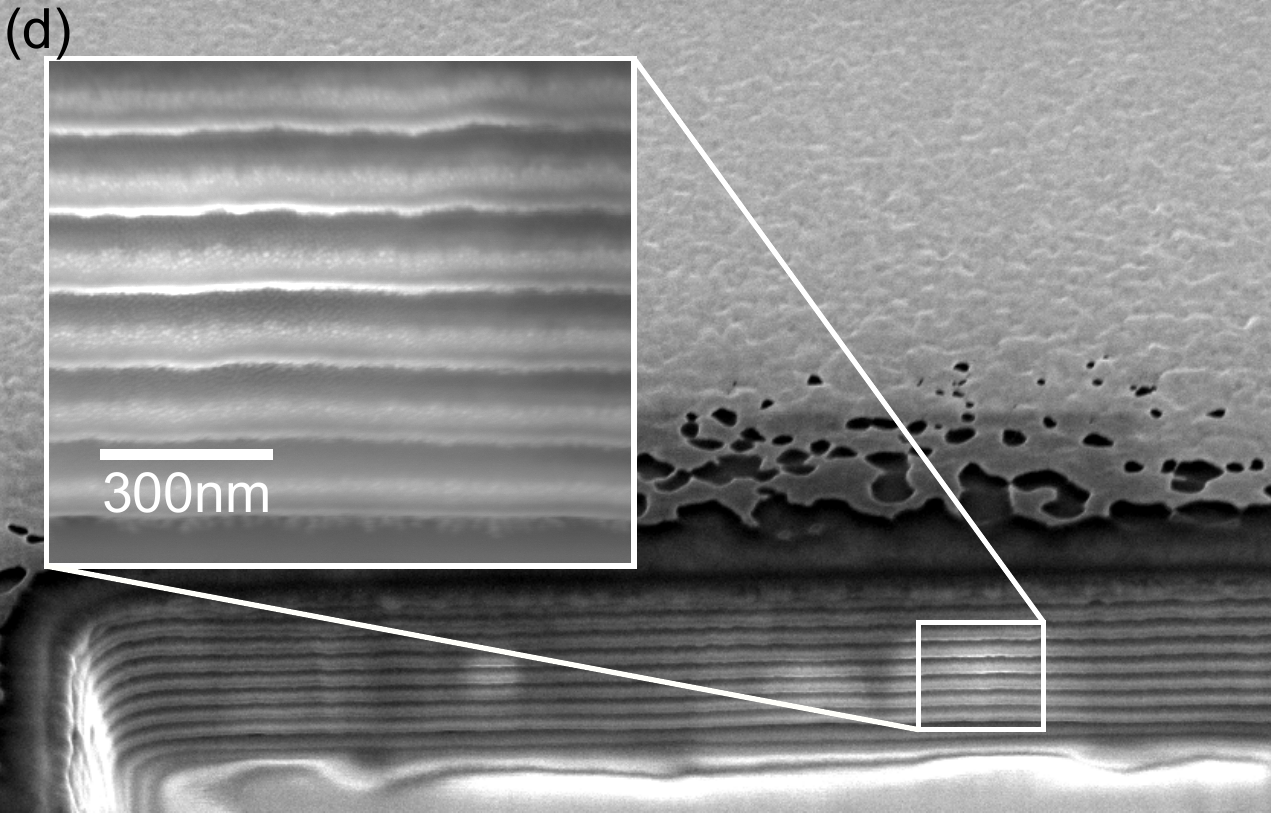}
  \includegraphics[width=0.32\linewidth,keepaspectratio,valign=t]{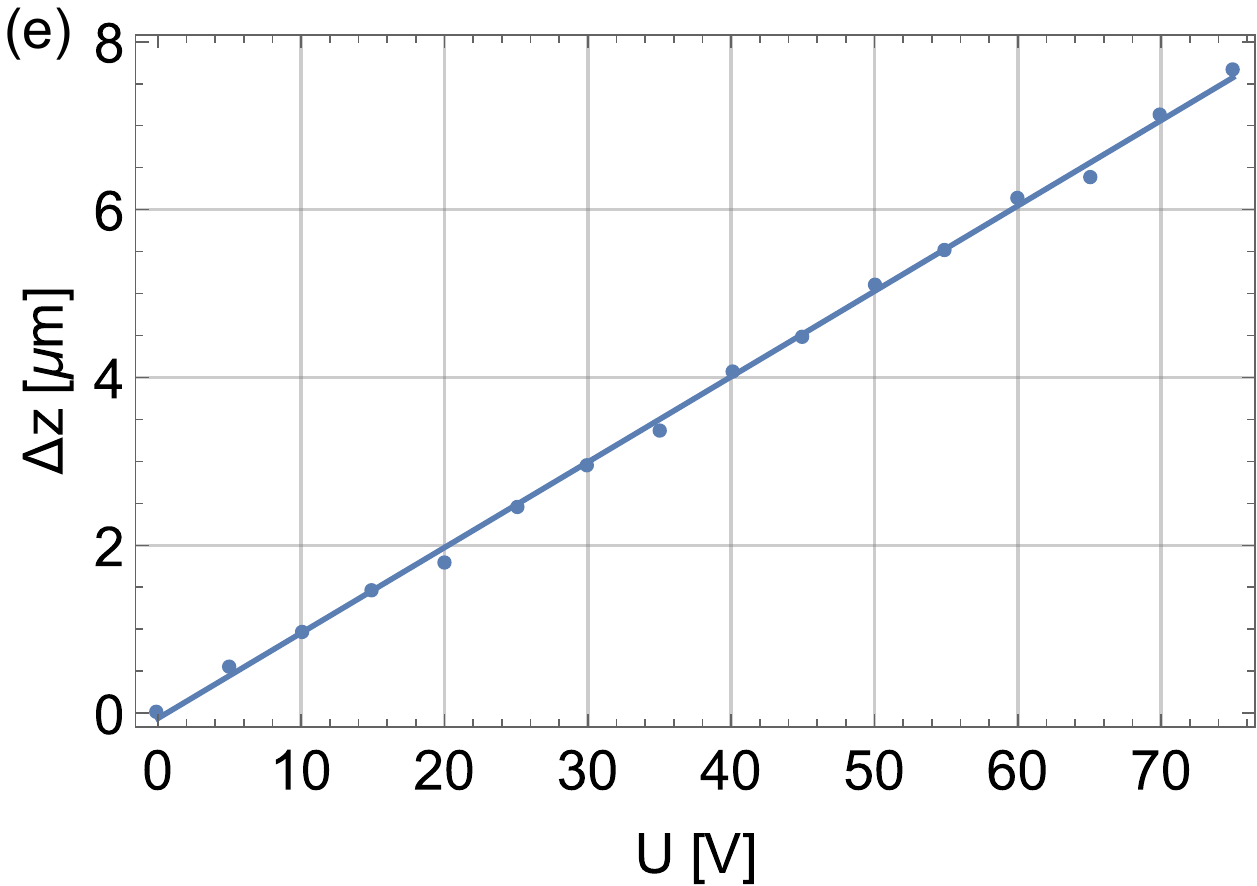}
  \includegraphics[width=0.32\linewidth,keepaspectratio,valign=t]{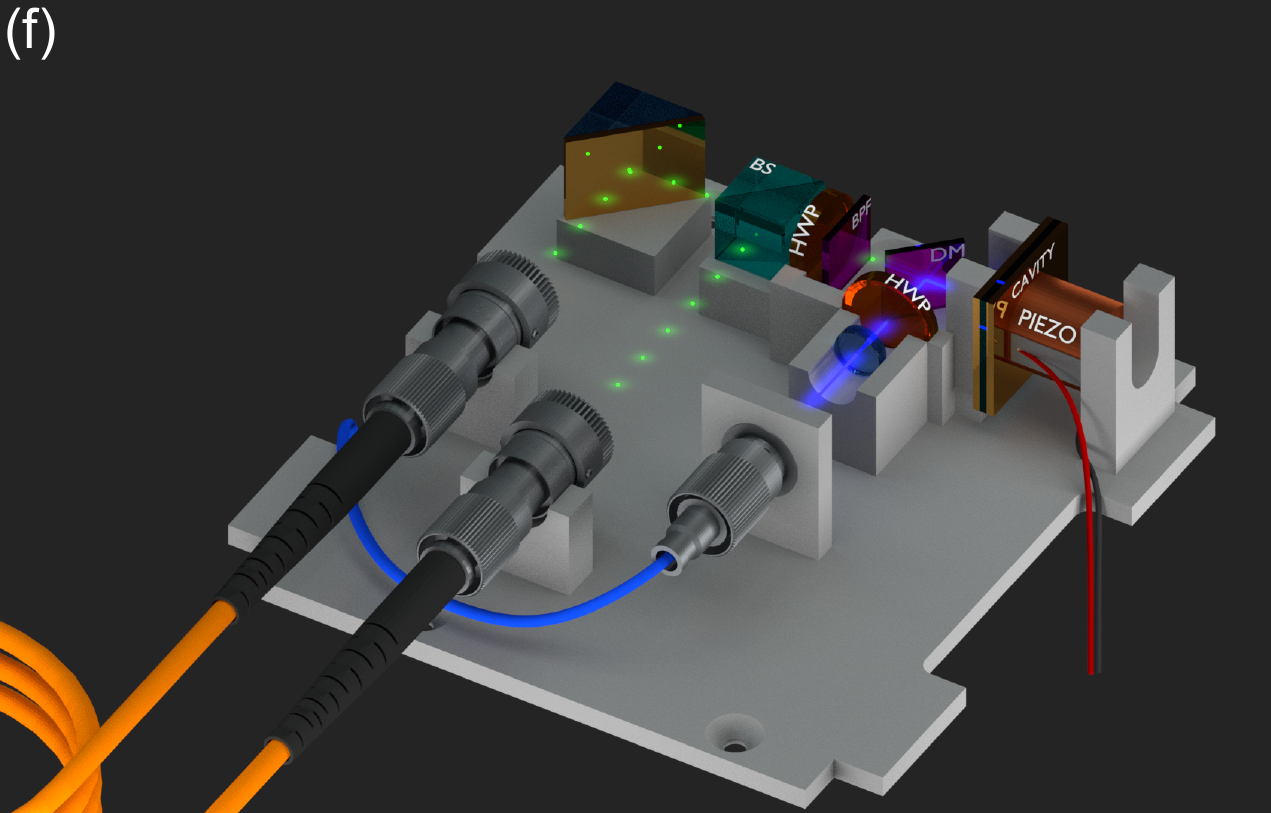}
\caption{Design and fabrication. (a) The microcavity consists of a hemispherical and flat mirror (only two stacks shown on either sides). The quantum emitter hosted by hBN emits confocally with the excitation laser. A PDMS spacer sets the cavity length. To prevent influence of the polymer on the emitter, the PDMS is etched in the middle. (b) Microscope image of the array of hemispheres (not all 64 shown). The surface profile of the hemisphere actually used for the cavity is shown in the right inset. The bottom inset shows the height profile through an arbitrarily chosen axis. The solid blue line shows an ideal cross section of a hemisphere with radius 2.7$\,\mu$m. (c) Reflectivity of the coating measured by spectrophotometry, with $R=99.2\%$ at the target wavelength $\lambda=565\,$nm. The inset shows the calculated cavity reflectivity based on the coating. (d) SEM image (immersion mode) of the mirror stacks, coated with a layer of gold. The sample is tilted by $52^\circ$, so the image is skewed in the vertical direction. The lighter areas in the cross section are regions which have been imaged with a magnification of $125000\times$ (see inset). The intense electron beam makes the surface reactive, and carbon-contaminations by residual organic materials in the SEM chamber are bonded at these areas. (e) Thickness change of a PDMS film with driving voltage reveals linear tuning with 102$\,$nm$\cdot$V$^{-1}$. (f) Design of the CubeSat platform (all components to scale). A polarization maintaining fiber (blue) guides the excitation laser from the diode below the platform. The laser is focused to the diffraction limit into the cavity onto the defect. The single-photons transmit through the dicroic mirror and are additionally band-pass filtered. Next, they are split by a 50:50 beam splitter and fiber-coupled into multimode fibers.}
\label{fig:1}
\end{figure*}

The confocal microcavity consists of a hemispherical and a flat mirror, with the hBN flake hosting the quantum emitter transferred to the focal point of the cavity (see Figure \ref{fig:1}(a)). The hemisphere spatially confines the cavity mode to the location of the emitter and is fabricated using I$_2$-enhanced focused ion beam milling\cite{10.1364/OL.35.003556,10.1364/OE.23.017205}. We fabricated arrays of 64 hemispheres per substrate with varying geometrical parameters. The surface roughness could be minimized by adding I$_2$-gas during the milling process. With the FIB we can achieve radii of curvature down to $<3\,\mu$m (see Figure \ref{fig:1}(b)). We initially characterized the hemispheres using an atomic force microscope (AFM) and phase-shift interferometry (PSI). The characteristic parameters extracted with both methods agree well, which allows us to use the much faster PSI for the characterizations. The hemisphere profile shown in Figure \ref{fig:1}(b) has a radius of 2.7$\,\mu$m and root mean square deviations $<1\,$nm from an ideal hemisphere\cite{supp_mat}. Note that we did not fabricate full hemispheres and the shapes deviate at the edges (which is due to a conductive coating to prevent charging effects during the milling). Both the flat and concave substrate are coated with 9 pairs of alternating dielectric quarter wave stacks (SiO$_2$/TiO$_2$), deposited using plasma sputtering. We measured a reflectivity of 99.2$\%$ at a wavelength of 565$\,$nm (see Figure \ref{fig:1}(c)). The calculated resulting cavity reflectivity (see Figure \ref{fig:1}(c), small inset) has a FWHM of 0.169$\,$nm, corresponding to a quality factor of $Q=3345$. The stopband of the cavity requires the single-photon excitation laser to be shorter than 504$\,$nm, otherwise the cavity has to be resonant at both the ZPL and excitation wavelength. By cutting through one of the stacks with a FIB and imaging with a scanning electron microscope (SEM) in immersion mode (see Figure \ref{fig:1}(d)), we see that stacking defects occur, as expected predominantly in higher layers. This is not an issue, however, as they are still $\ll\lambda$. The reflectivity is most likely limited by incorporated residual nitrogen, leading to scattering losses. It should also be noted that at a high magnification (see inset) the stacks show a spotted pattern. This is actually re-deposition of atoms during the milling with the FIB. The backsides of the substrates were coated with anti-reflective coatings, consisting of a single quarter wave layer MgF$_2$. This reduces the reflection losses at the glass-air interface from 4.33$\%$ to 2.97$\%$.\\
\indent Multilayer hBN flakes have been placed onto the flat mirror via clean polymer transfer (see Methods). The more common direct dry transfer was not used as this usually also transfers residues. The hBN crystals were treated using an oxygen plasma followed by rapid thermal annealing under an Ar atmosphere\cite{10.1021/acsphotonics.8b00127}. Using plasma etching, defects with their ZPL primarily around 560$\,$nm form, well within the stopband of the coating. Finally, a tuneable polymer spacer is deposited onto the concave mirror. A piezoelectric actuator provides the tuning force and compresses the polymer. In contrast to monolithic cavities\cite{10.1038/nphoton.2014.304,10.1038/ncomms13328,10.1364/OME.9.000598}, this approach allows for \textit{in-situ} tuning of the cavity length. The tuning capability is essential, since the exact position of the ZPL cannot yet be controlled and the optical cavity mode has to be artificially matched to the spectrum of the emitter\cite{10.1038/s41467-018-05117-4}. Due to a suitable Young's modulus and the ability to deform reversibly, we selected PDMS (polydimethylsiloxane) from a range of polymers\cite{supp_mat}. Figure \ref{fig:1}(e) shows that the compression of the PDMS film is linear with the driving voltage at the actuator, with a tuning of 102$\,$nm$\cdot$V$^{-1}$. This allows us to easily lock the cavity to any arbitrary wavelength. To prevent influence of the PDMS on the emitter, the PDMS was etched around the array.\\
\indent The cavity mirrors, together with all in- and out-coupling optics, were aligned and glued to a monolithic platform (see Figure \ref{fig:1}(f) and \cite{supp_mat}). Prior to the gluing each component, held with vacuum tweezers, has been aligned with a motorized 6-axis translation stage. This greatly reduces the size of the complete SPS, at the cost of limiting the tuneability to only cavity length. Changing the radius of curvature of the cavity as demonstrated in a similar experiment is thus not possible\cite{10.1063/1.5026779}. Nevertheless, the compact size of optics, as well as choice of electronics and excitation laser, allow us to reduce the size of the full experiment to $10\times 10\times 10\,$cm$^3$. This is the size requirement of the 1U CubeSat standard, a miniature pico-class satellite. This makes the single-photon source portable and a promising candidate for low cost CubeSat-based single-photon QKD, especially as the quantum ermitters in hBN are space-certified\cite{arXiv:1811.10138}.

\section{Performance of the single-photon source}
\begin{figure*}[t!]
\centering
  \includegraphics[width=0.32\linewidth,keepaspectratio,valign=t]{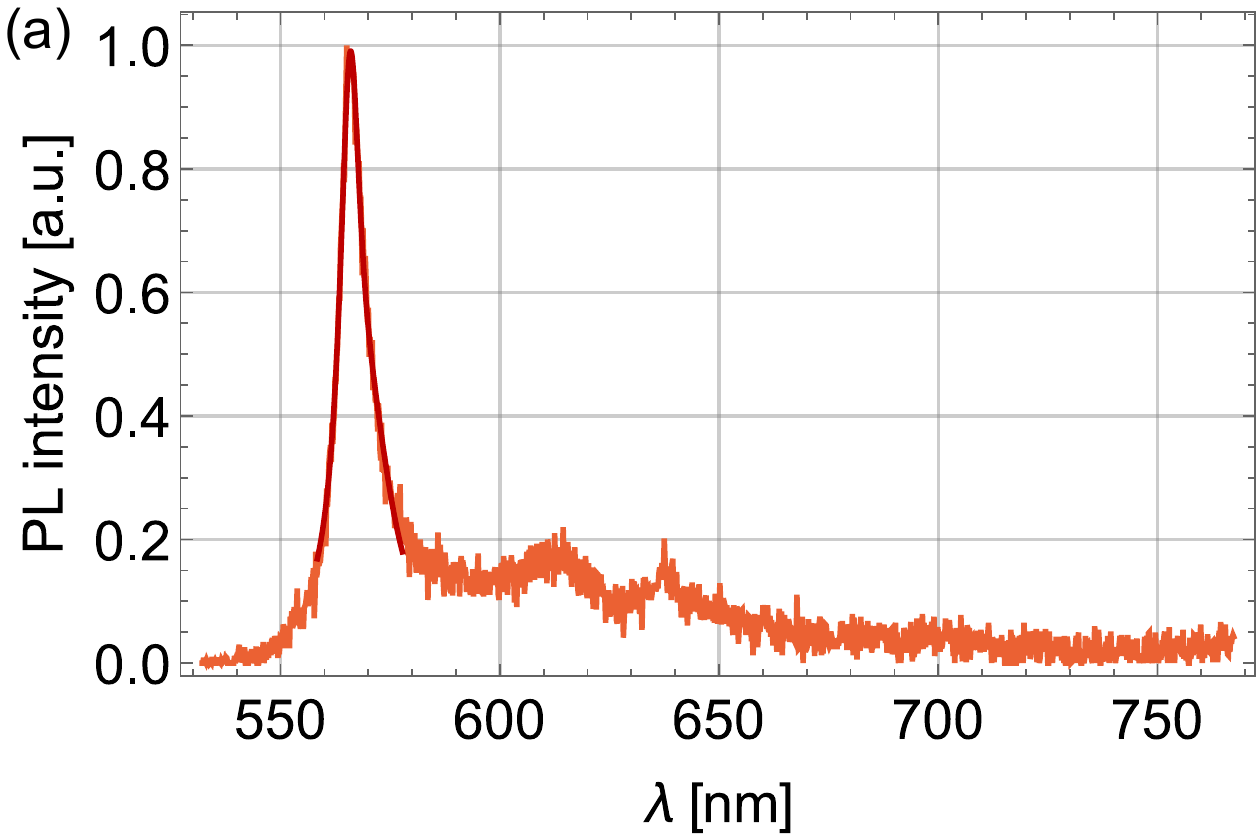}
  \includegraphics[width=0.32\linewidth,keepaspectratio,valign=t]{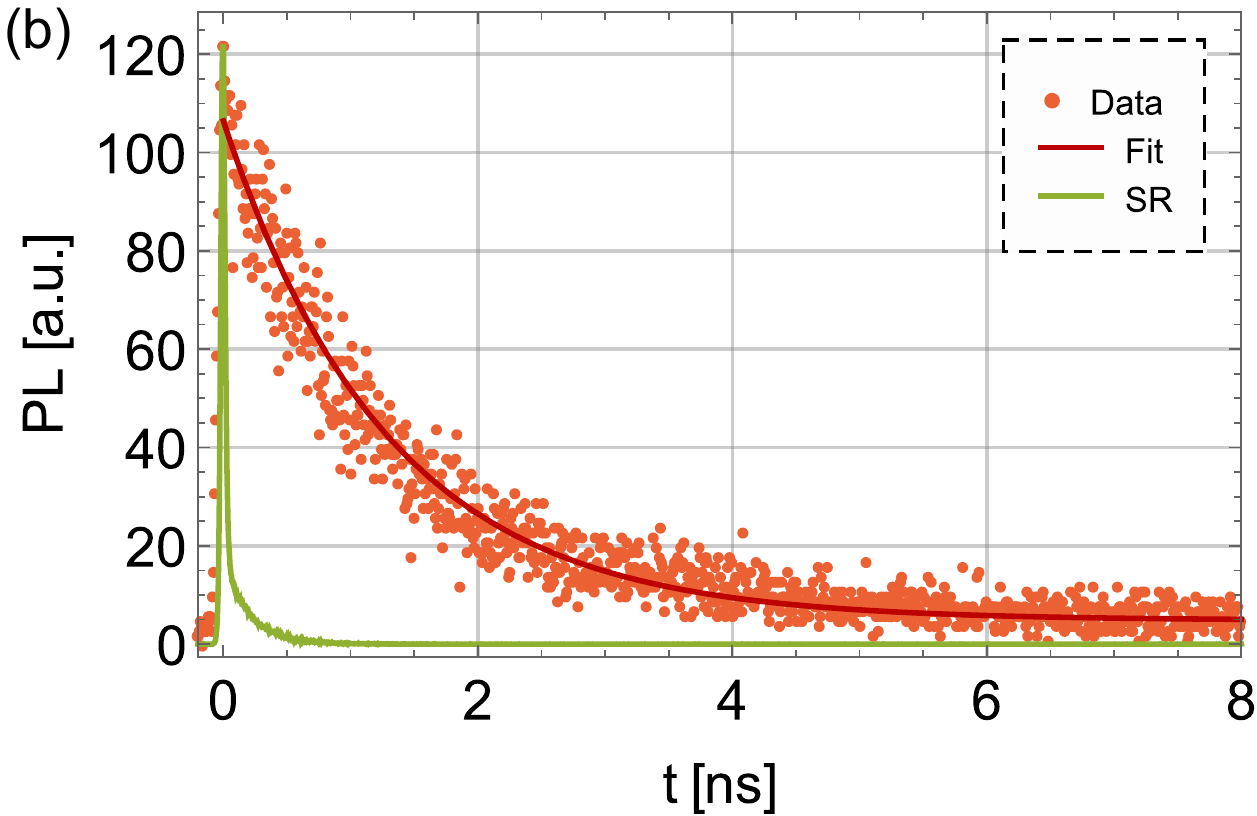}
  \includegraphics[width=0.32\linewidth,keepaspectratio,valign=t]{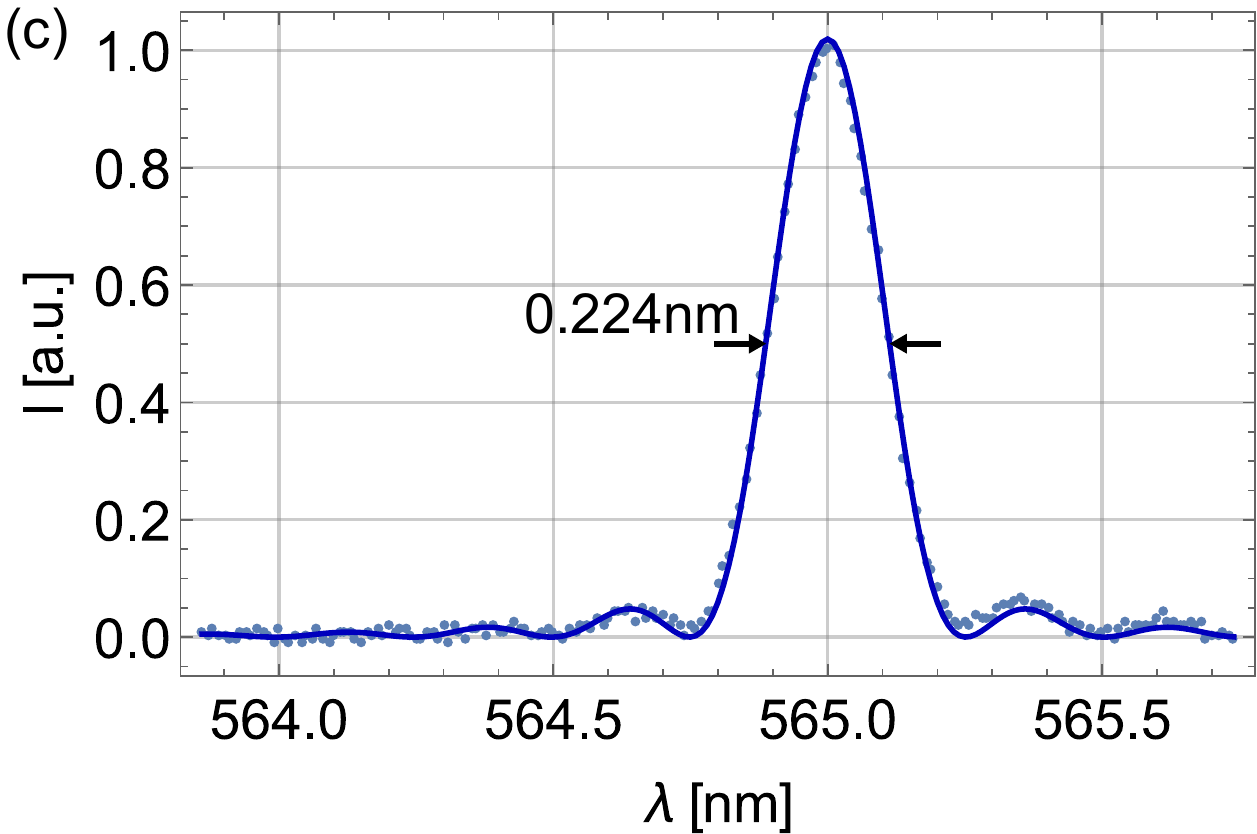}\\
  \vspace{0.1cm}
  \includegraphics[width=0.32\linewidth,keepaspectratio,valign=t]{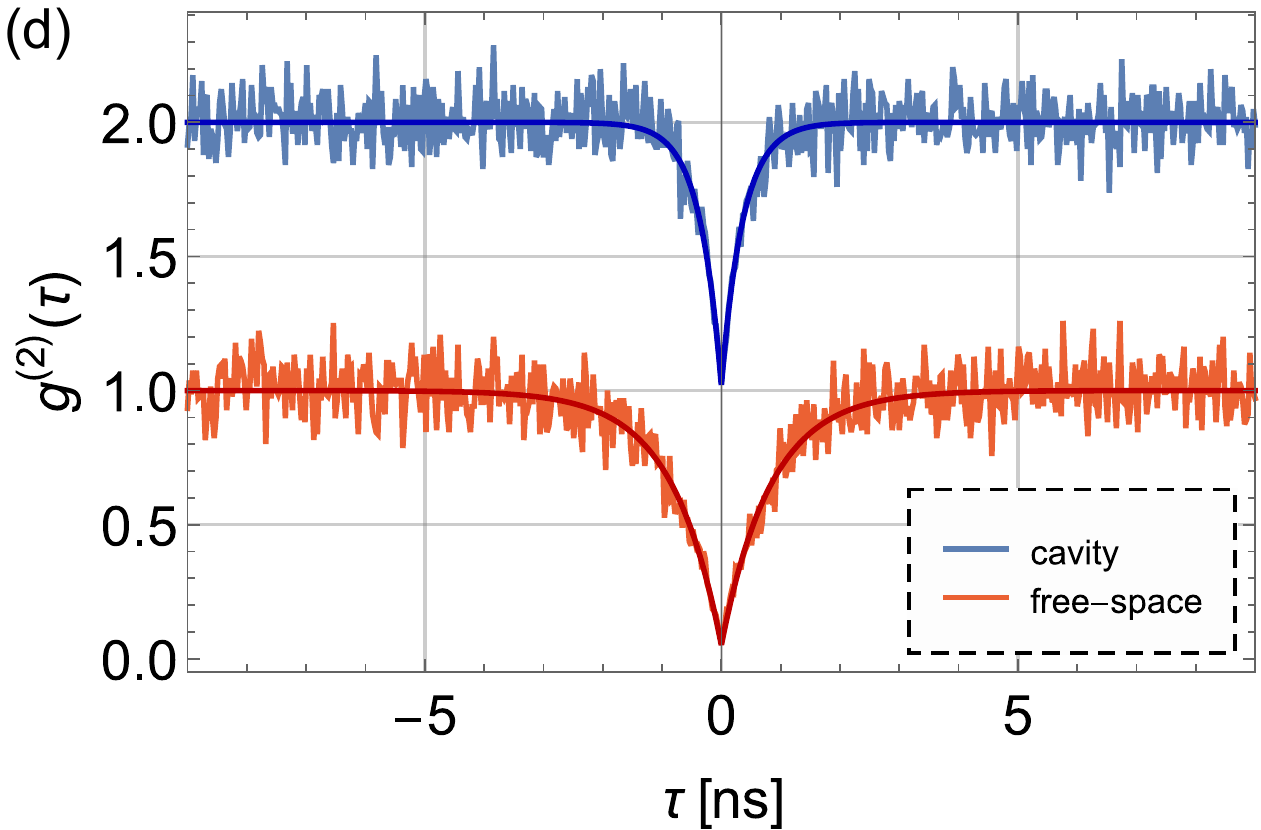}
  \includegraphics[width=0.32\linewidth,keepaspectratio,valign=t]{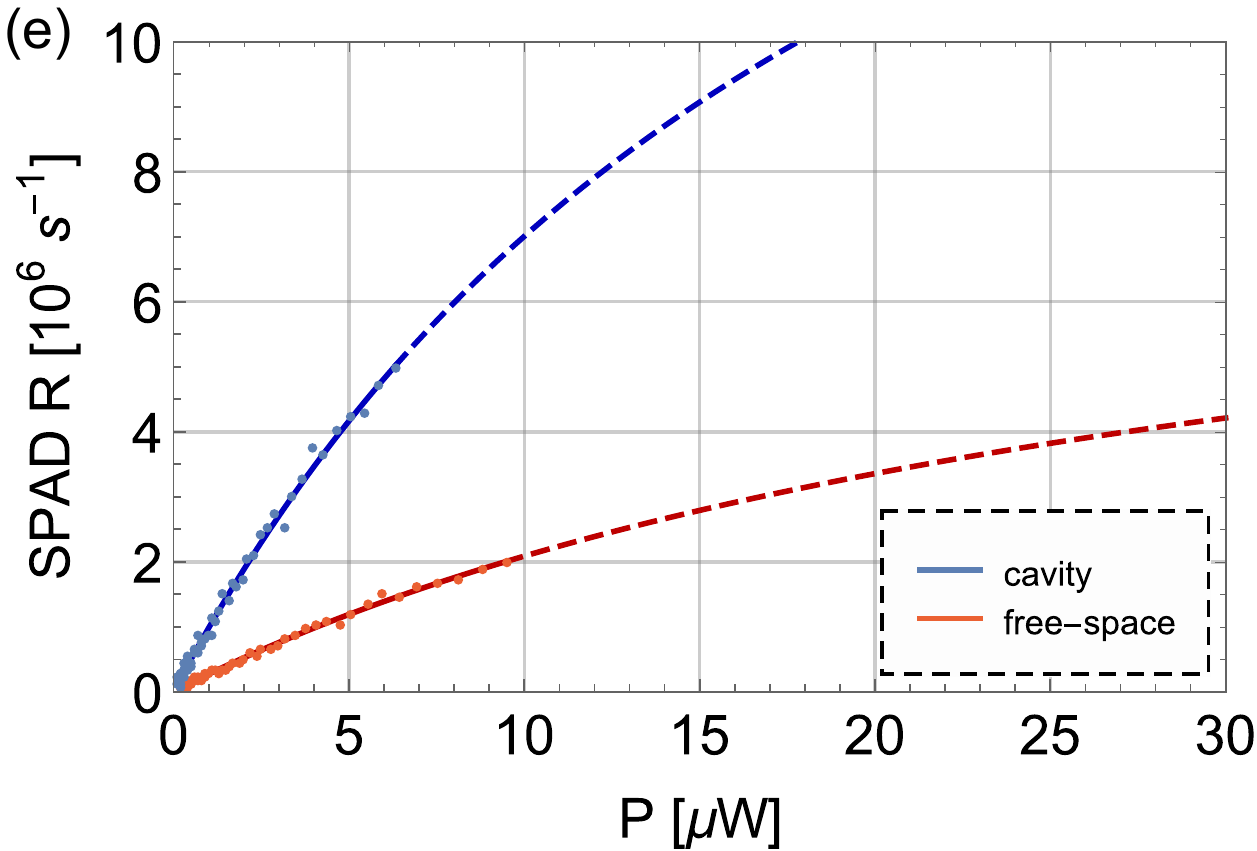}
  \includegraphics[width=0.32\linewidth,keepaspectratio,valign=t]{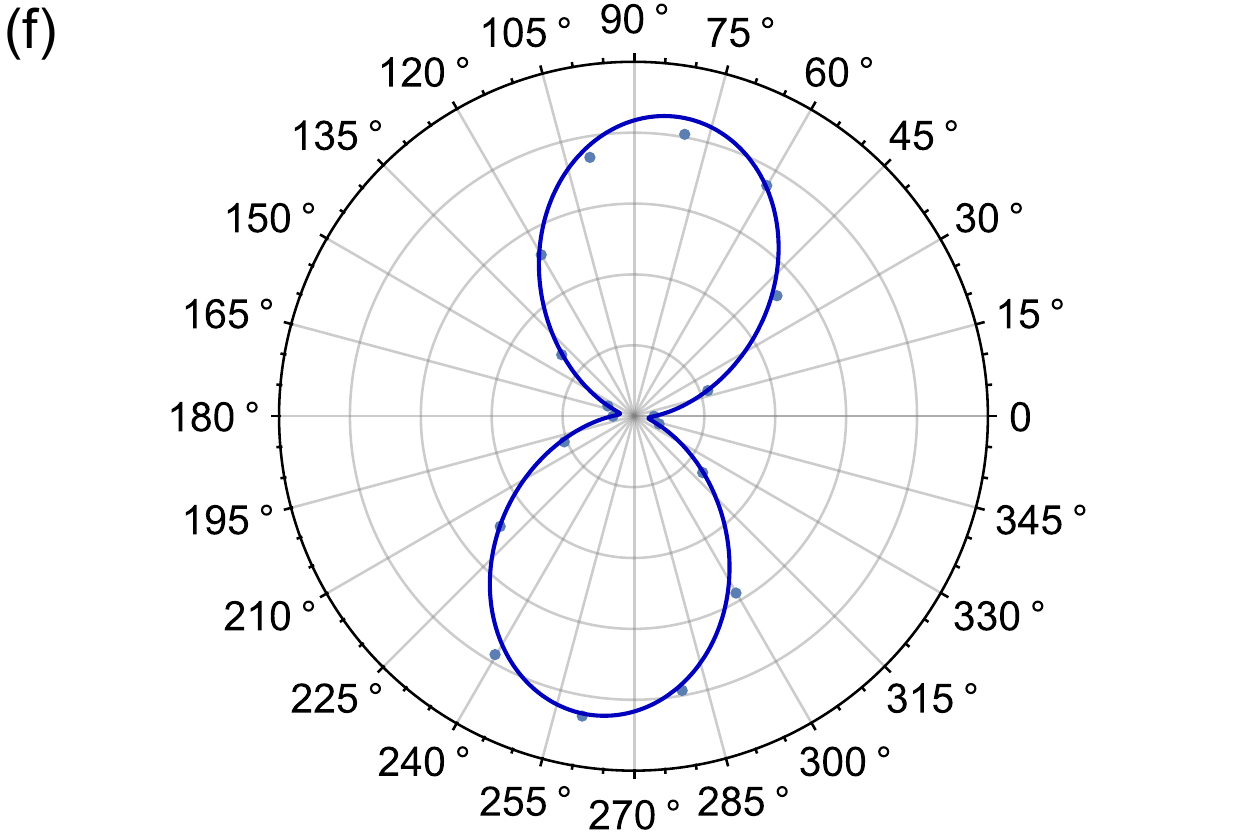}
\caption{Performance of the single-photon source. (a) Free-space spectrum after off-resonant excitation measured in-reflection and coupled to a grating-based spectrometer. From a Lorentzian fit we extract the ZPL at 565.85(5)$\,$nm and a linewidth (FWHM) of 5.76(34)$\,$nm. (b) Time-resolved photoluminescence reveals an excited state lifetime of $\tau=897(8)\,$ps. The exponential fit function is convoluted with the system response (SR). (c) The cavity narrows the spectrum down to 0.224$\,$nm (FWHM). The spectrum has been recorded using a high-resolution Fourier-transform spectrometer. The finite scan range result in the spectrum being convoluted with the system response function (of the form of a $\sinc^2(x)$), which in turn leads to the side lobes. (d) When comparing free-space with cavity-coupled emission, the second-order correlation function measurements show a decrease of $g^2_0$ from 0.051(23) or -12.9$\,$dB to 0.018(36) to -17.4$\,$dB and shortening of the lifetime from 837(30) to 366(19)$\,$ps due to the Purcell effect. The cavity data is vertically offset for clarity. (e) The cavity increases the single-photon count rate, even at lower excitation power. This is because of the shortening of the lifetime due to the Purcell effect, but also due to an enhanced collection efficiency with the cavity. (f) The emission is dipole-like, as the projections on different polarization directions show. The solid line is obtained by fitting a $\cos^2(\theta)$ function.}
\label{fig:2}
\end{figure*}

Prior to the cavity experiments we performed a free-space characterization of the quantum emitter on the mirror. All measurements were carried out at room temperature. The defects were located using confocal photoluminescence (PL) mapping under off-resonant excitation at 522$\,$nm. As hBN itself is optically inactive in the visible spectrum, all emission originates from the defects or surface contaminants. Each crystal is scanned with a resolution of 0.5$\,\mu$m. For the cavity, we selected a suitable defect with a ZPL at 565.85(5)$\,$nm and a Lorentzian linewidth (FWHM) of 5.76(34)$\,$nm (see Figure \ref{fig:2}(a)). The PL spectrum shows the typical asymmetric lineshape. Note that this is not a result of partial suppression of the long pass filter used to block the pump laser (see Methods), but rather the PSB being adjacent to the ZPL. The defect emits 63.2$\%$ into its ZPL. We note that the emission $>580\,$nm originates from surface contaminants activated during the annealing and is usually filtered out (see Methods). Alternatively, annealing in a reactive environment can burn off these contaminants. Time-resolved PL (TRPL) reveals a single-exponential decay with a lifetime of 897(8)$\,$ps (see Figure \ref{fig:2}(b)). The fit function is convoluted with the system response (also shown in Figure \ref{fig:2}(b)) in order to reproduce the observed data.\\
\indent For the cavity experiments we used a custom-built high-resolution Fourier-transform spectrometer (FTS), instead of the grating-based spectrometer. After aligning the concave to the flat mirror and coupling the emitter with the cavity mode, we saw an improved spectral purity (see Figure \ref{fig:2}(c)), with the single-photon linewidth narrowing down to 0.224$\,$nm (FWHM). In frequency space this corresponds to 210.6$\,$GHz. The spectrum, however, shows side lobes, which do not originate from higher-order transverse cavity modes. The transverse mode spacing is much larger than the difference in observed peak positions\cite{supp_mat}. These peaks are artifacts from the finite scan range of the FTS which results in a truncated Fourier-transform. Convoluting this response (which is of the form of $\sinc^2(x)$) with a Lorentzian reproduces the observed data.\\
\indent The lifetime cannot be measured directly using time-resolved PL, as the wavelength of the ultra-short pulsed laser is within the stopband of the cavity. For a single-photon emitter, however, it is possible to extract the lifetime directly from the second-order correlation function, which we measure using a Hanbury Brown and Twiss (HBT)-type interferometer. For the emitter in the cavity, we measure $g^2_0\equiv g^{(2)}(\tau=0)=0.018(36)$ (see Figure \ref{fig:2}(d)) and from the fit we extract a lifetime of 366(19)$\,$ps (see Methods). For a fair comparison of free-space and cavity-enhanced lifetimes we compare the correlation function measurements in free-space and with the cavity. The $g^{(2)}(\tau)$ for the uncoupled emitter dips only to 0.051(23) and has a lifetime of 837(30)$\,$ps. The lifetimes measured with time-resolved PL and extracted from the $g^{(2)}(\tau)$ measurements agree reasonably well, even though we note that the 897(8)$\,$ps from the TRPL measurement is likely more accurate. While $g^2_0=0$ is within the error margin for the cavity-coupled emitter, more accurate measurements are required to reduce the error margin to extract the true value of $g^2_0$. Knowing this is crucial for QKD applications (see below). A small error margin on correlation function measurements can typically be achieved with ultra-short pulsed excitation\cite{10.1038/s41534-018-0092-0,10.1063/1.5020038}. We also calculated the background correction term\cite{PhysRevA.64.061802} and found that it is smaller than the significant digits of our measurement result ($<5\times 10^{-5}$), so we conclude that any deviation from 0 is not due to detector dark counts, but rather other noise sources excited through the laser. If we directly compare $g^2_0$ of the uncoupled and cavity-enhanced emitter, however, we see a reduction of a factor of 2.83. Such reduction can typically be achieved in the "bad-emitter" regime and means that off-resonant noise sources are successfully suppressed. A narrower cavity linewidth could thus further reduce $g^2_0$. The ratio of free-space (or rather half-sided cavity) to cavity-coupled lifetime is $f=2.29$. The effective Purcell enhancement is given by
\begin{align}
F_p^{\text{eff}}=\frac{3}{4\pi^2}\lambda^3\frac{Q^{\text{eff}}}{V}
\end{align}
with $Q^{\text{eff}}$ being the effective quality factor and $V$ being the cavity mode volume. We calculate the mode volume to be 1.76$\lambda^3$. In the "bad-emitter" regime the effective quality factor $Q^{\text{eff}}=\frac{\lambda}{\Delta\lambda_{\text{cav}}+\Delta\lambda_{\text{em}}}$ has to be used, which is dominated by the emitter dynamics. It should be mentioned that this is only an approximation and it is more accurate to calculate the overlap integral of the photonic density of states of the cavity and electronic density of states of the emitter. In addition, this effective Purcell factor is different from the ratio $f$, because the dielectric environment of the mirror is modifying the available density of states, whereas the Purcell factor is the ratio of vacuum (or true free-space) to cavity lifetime. We calculate the effective Purcell factor to be 4.07. This also allows for the direct calculation of the quantum efficiency\cite{10.1063/1.5026779}, given by
\begin{align}
\eta=\frac{f-1}{f+F_p^{\text{eff}}-\varepsilon f}
\end{align}
where $\varepsilon$ is the Purcell factor caused by the mirror and is determined by finite-difference time-domain (FDTD) simulations. For our mirror we find $\varepsilon=1.68$\cite{supp_mat} and thus the quantum efficiency is 51.3$\%$.\\
\indent The cavity also modifies the power saturation behavior (see Figure \ref{fig:2}(e)), with an increased single-photon count rate even at lower excitation power. This is a result of the Purcell enhancement, which makes the emitter brighter, but also from the increased collection efficiency of the cavity, as the emitter predominantly emits into the cavity mode. The low excitation power also assists the single-photon count rate stability, because at low excitation power the emitters show no blinking or photobleaching. This is particular important as the photobleaching increases with decreasing wavelength\cite{10.1021/acsphotonics.6b00736} and due to the stopband of the cavity our excitation laser is at 450$\,$nm. Note that the count rates at the single-photon avalanche diodes (SPADs) shown in Figure \ref{fig:2}(e) are the raw count rates, not corrected for transmission loss or detector efficiency. The quantum emitter also emits linearly polarized light (see Figure \ref{fig:2}(f)) with a degree of polarization (DOP) of 90.4$\%$. The fit is obtained using a $\cos^2(\theta)$ function. A high polarization contrast is crucial for QKD applications which use polarization encoding. Increasing the DOP of not fully polarized light is always accompanied by loss, and so it sets an upper bound on the efficiency of the SPS.\\
\indent Since the cavity length is tuneable, the single-photon wavelength can also be tuned. Effectively, the tuning range is the linewidth of the free-space emission. The cavity is only sampling the free-space emission spectrum, however, so the actual single-photon count rate is the spectral overlap integral of optical cavity mode and emitter. This results in the emission rate decreasing with increasing cavity detuning.

\section{Theoretical modeling}
\subsection{Numerical modeling}
We can use FDTD simulations to calculate the electric field distribution of the dipole emitter in the cavity. The electric field intensity ($|E|^2$) is shown in Figure \ref{fig:3}(a). The simulations also show that resonance does not occur at a physical mirror separation $L'$ which is a multiple of $\lambda/2$. This is due to the finite penetration depth of the electric field into the dielectric mirror stacks, leading to an effective cavity length. The penetration depth $\xi$ thereby is given by
\begin{align}
\xi=\frac{q\lambda/2-L'}{2}
\end{align}
The physical mirror separation $L'$ is determined by maximizing the intracavity electric field\cite{supp_mat}. Our simulations yield $\xi=122\,$nm. When designing the thickness of the PDMS spacer, this has to be taken into account. To reduce the computational time we simulated the longitudinal mode $q=5$ instead of the experimentally realized $q=8$. Nevertheless, the parameter $\xi$ is not affected by this beyond the numerical precision of the simulation.

\subsection{Applications in quantum technologies}
\begin{figure*}[t!]
\centering
  \includegraphics[width=0.32\linewidth,keepaspectratio,valign=t]{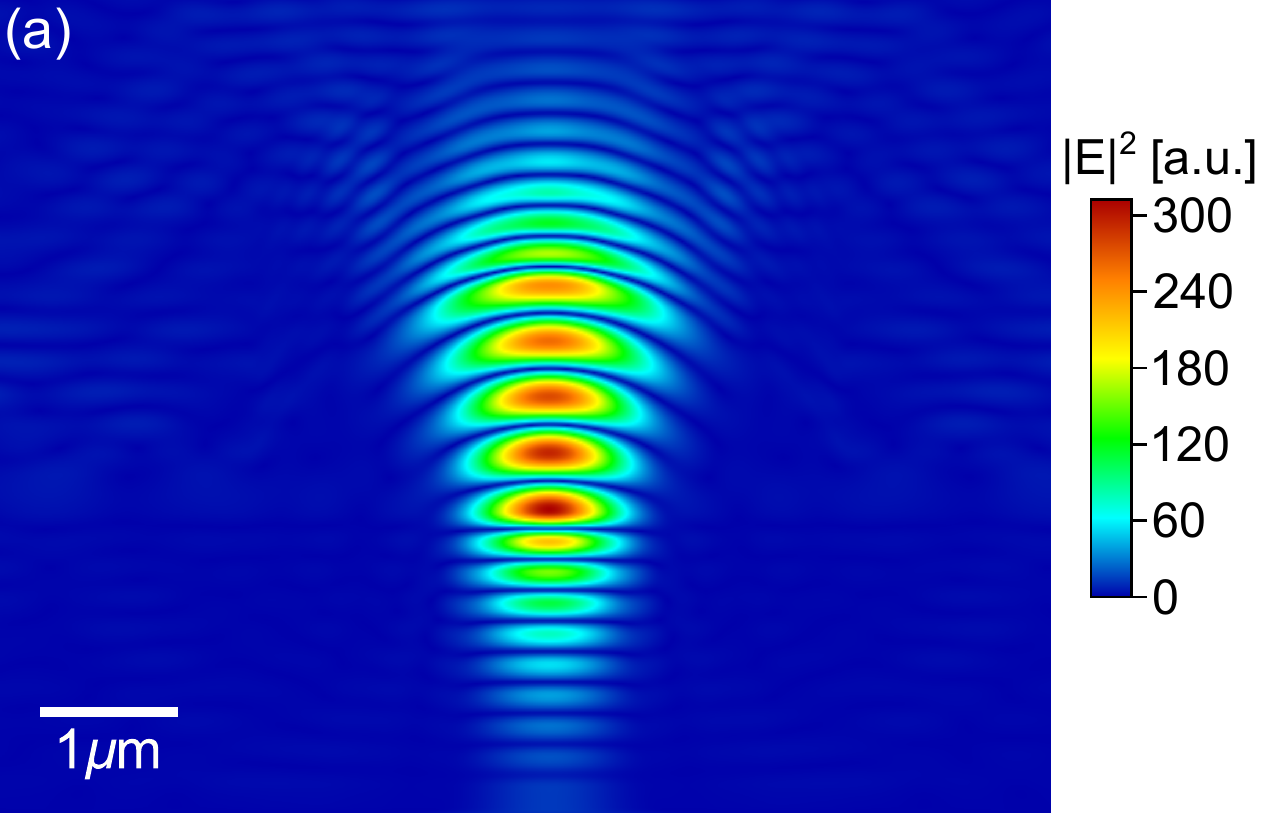}
  \includegraphics[width=0.32\linewidth,keepaspectratio,valign=t]{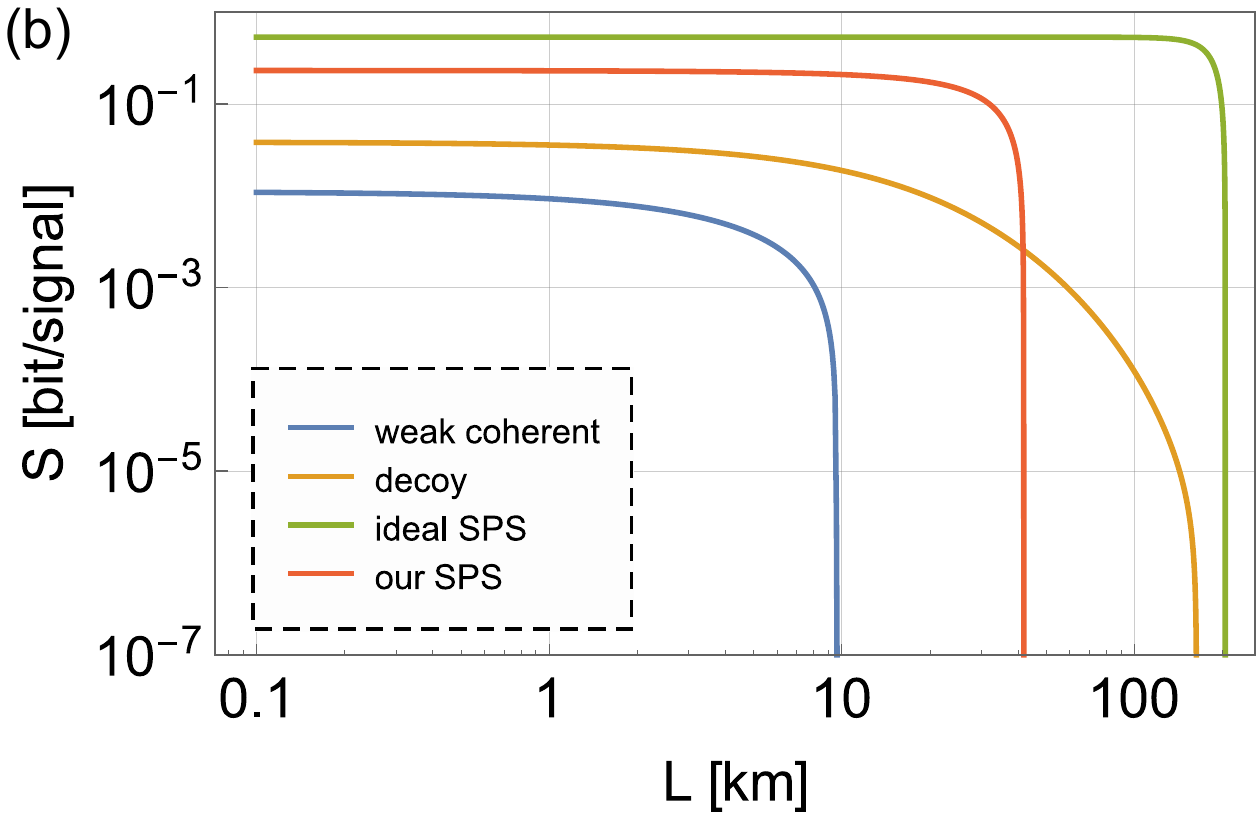}
  \includegraphics[width=0.32\linewidth,keepaspectratio,valign=t]{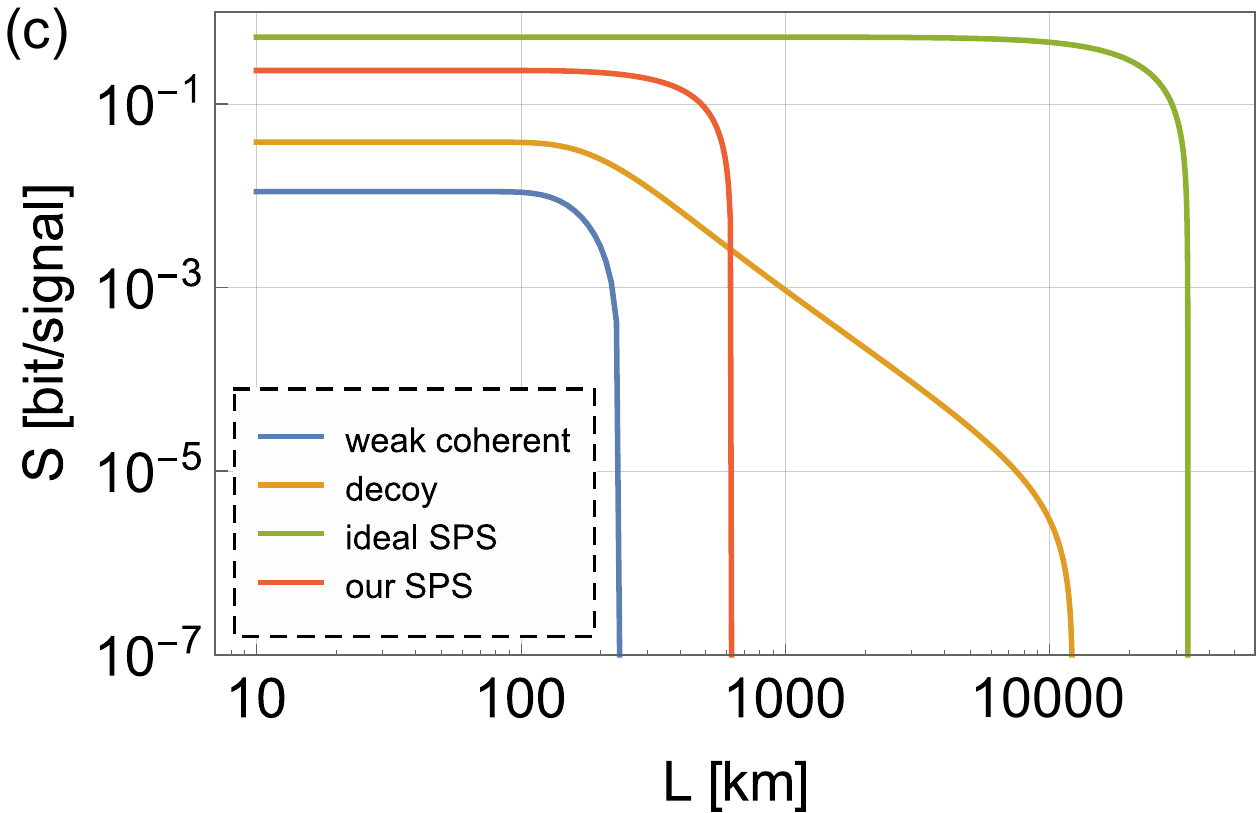}\\
  \vspace{0.1cm}
  \includegraphics[width=0.32\linewidth,keepaspectratio,valign=t]{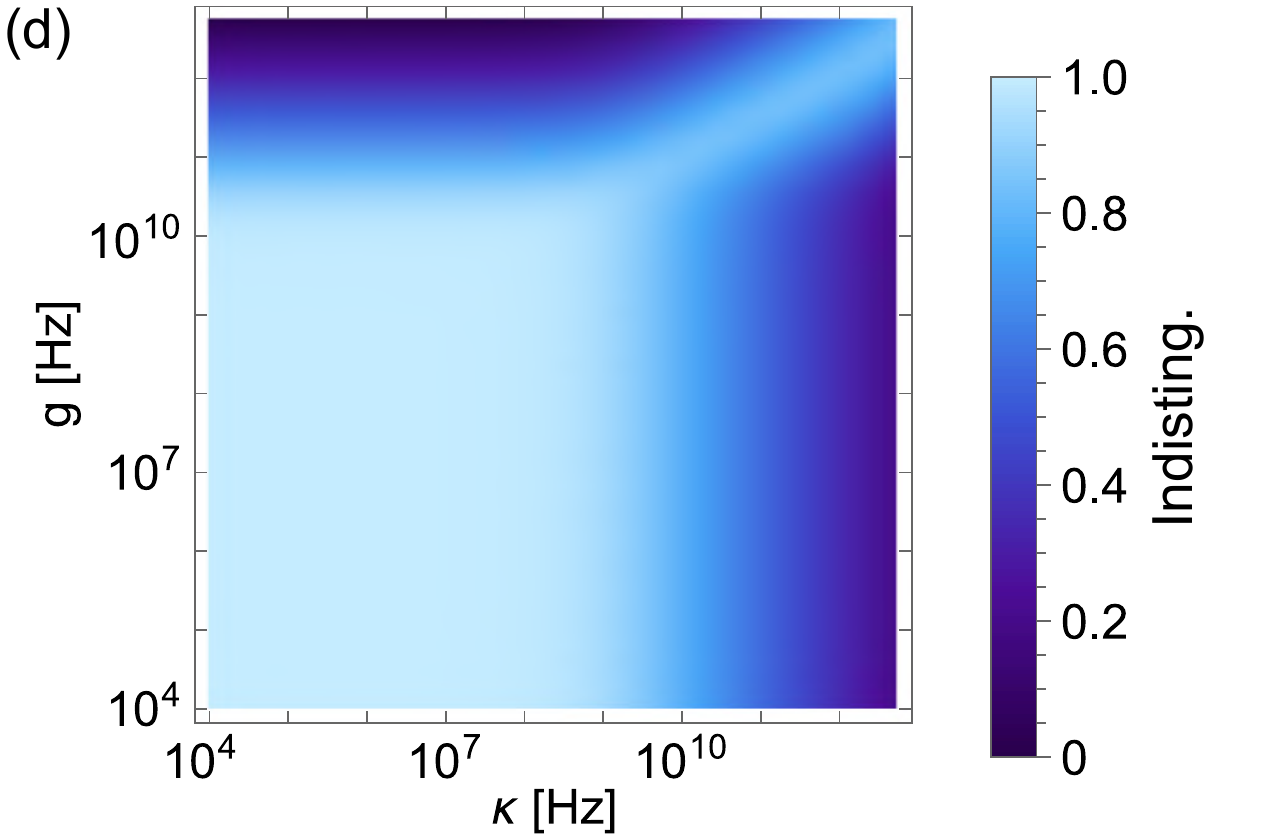}
  \includegraphics[width=0.32\linewidth,keepaspectratio,valign=t]{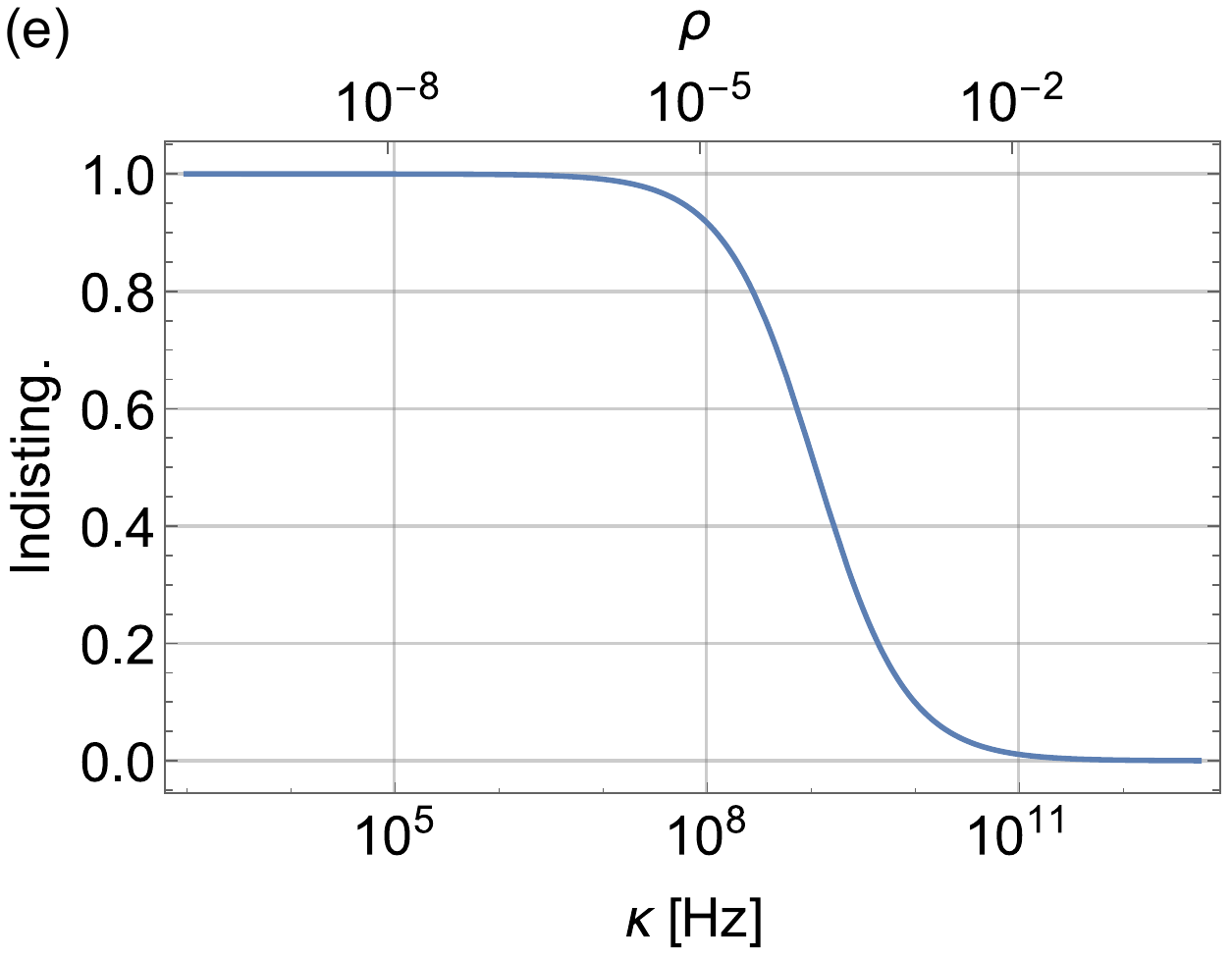}
\caption{Theoretical modeling. (a) Electric field mode profile of a dipole emitter in the cavity obtained using FDTD simulations. (b) QKD for different photon sources for a fiber channel with 0.21$\,$dB/km loss. Our SPS outperforms weak coherent states at all distances and decoy state at short to medium distances up to 42$\,$km. (c) QKD for different photon sources for a free-space satellite-to-ground link. The satellite assumes a 5$\,$cm telescope, the ground station a 60$\,$cm telescope. Our SPS outperforms decoy states at distances up to 630$\,$km and weak coherent states at all distances. For both channels, our single-photon source assumes $g^2_0=0.018$ and a quantum efficiency of 51.3$\%$\cite{supp_mat}. (d) Indistinguishability in the weak coupling limit as a function of cavity coupling rate $g$ and cavity linewidth $\kappa$. The simulations assume the photophysics of our actual emitter in the cavity. A high indistinguishability can be achieved for $g,\kappa<10^9\,$Hz. (e) Indistinguishability in the limit $g\ll 10^8\,$Hz. $I>0.9$ requires $\kappa<124\,$MHz.}
\label{fig:3}
\end{figure*}

We now turn to an evaluation of the SPS for the two most common quantum information applications: quantum key distribution and quantum computing. Due to the lack of suitable SPSs the vast majority of QKD implementations use weak coherent states (WCSs). These are characterized by a low mean photon number (resulting in a low efficiency of the protocol) and a non-zero probability of emitting two or more photons at the same time. The multi-photon pulses contain information leaking to a potential eavesdropper. This can be reduced at the expense of sacrificing parts of the exchanged key. In comparison, an ideal single-photon source has an efficiency of unity and no multi-photon emission, so it always performs better than any protocol based on WCSs. We assess the performance of our single-photon source for the BB84 protocol\cite{10.1103/RevModPhys.74.145} over a fiber channel with a loss of 0.21$\,$dB/km and realistic parameters from the experiment by Gobby, Yuan and Shields (GYS)\cite{10.1063/1.1738173}. It should be mentioned that such loss can only be achieved at telecom wavelengths, where single-photon emission from hBN has yet to be demonstrated, but for simplicity we still use all GYS parameters. The relevant metric is the extractable secret bit per sent signal\cite{supp_mat}. Moreover, we compare the results with an ideal SPS and the most common conventional QKD protocols: weak coherent and decoy states\cite{PhysRevLett.94.230504}. The latter is the most efficient protocol that is publicly known. The simulations for weak coherent and decoy states assume a fixed mean photon number per pulse $\mu$, however, there is an optimal choice of $\mu$ for every distance.\\
\indent At short and medium distances below 42$\,$km, our SPS outperforms both weak coherent and decoy states, while at long distances decoy states become more efficient (see Figure \ref{fig:3}(b)). This is due to the fact that at long distances (i.e. high losses), multi-photon pulses harm the extractable secret bit disproportionately. Nevertheless, our SPS performs better than weak coherent states in each instance. Decoy state protocols can still achieve a finite secret key rate at large distances, because they can extract information from multi-photon states while at the same time defeating the photon-number splitting attack (multi-photon states dominate at long distances with high losses). A communication distance of $<42\,$km would be typical for metropolitan networks. The ideal SPS of course outperforms all protocols and also our source at all distances. For space-to-ground links the loss is dominated by diffraction and atmospheric attenuation plays a role only in the lowest 10$\,$km. The simulations (see Figure \ref{fig:3}(c)) show that our source outperforms the decoy state protocol on distances up to 630$\,$km. For a comparison: The Micius satellite, which performed the first satellite-to-ground quantum key exchange, orbits at around 500$\,$km\cite{10.1038/nature23655}. Thus, our single-photon source could enhance the key generation rate even with its current performance for such a satellite. We note that the free-space loss channel assumes only diffraction losses and no other noise sources such as pointing errors, atmospheric loss and losses in the transmitter or receiver, which would change the result only marginally. The crossing point where our SPS and the decoy state protocol perform equally efficient for both channels is at a loss of 8.82$\,$dB.\\
\indent Notably, QKD only requires maximal entropy on all degrees of freedom which are not used for qubit encoding. Other quantum information protocols, however, do require truly indistinguishable single-photons. An example are entangling gates for single-photons for use in one-way quantum computing\cite{PhysRevLett.86.5188}. A measure of how indistinguishable consecutively emitted single-photons are is the interference contrast $I$ in a Hong-Ou-Mandel experiment\cite{PhysRevLett.59.2044}. Unfortunately, as our cavity is pumped continuously, we cannot directly measure $I$. Nevertheless, we can at least theoretically calculate the expected indistinguishability. The indistinguishability of a quantum emitter with pure dephasing is given by
\begin{align}
I=\frac{\gamma}{\gamma+\gamma^{*}}
\end{align}
where $\gamma$ is the emission rate and $\gamma^{*}$ is the pure dephasing rate. At room temperature we find $I=2\times 10^{-4}$, meaning only 1 in 5000 photons would interfere in a Hong-Ou-Mandel experiment. Even such a strongly dephasing emitter, however, can reach a regime of high indistinguishability, when coupled with a high-Q cavity. In the limit of weak coupling $I$ modifies to
\begin{align}
I=\frac{\gamma+\kappa R/(\kappa + R)}{\gamma + \kappa + 2R}
\end{align}
where the parameter $R=\frac{4g^2}{\kappa+\gamma+\gamma^{*}}$ is the effective transfer rate between the emitter and the cavity, $\kappa$ is the cavity linewidth and $g$ is the cavity coupling strength\cite{PhysRevLett.114.193601}. For our cavity parameters we find $I=5.3\times 10^{-3}$. While this is an improvement by a factor of 26, it is still far beyond being useful for fault-tolerant quantum computing. The indistinguishability for generalized cavity linewidth and coupling strength is shown in Figure \ref{fig:3}(d). Note that in the limit of strong coupling it is also possible to achieve a high indistinguishability. With the coupling strength typically $\ll 1\,$GHz, a narrow cavity linewidth is required to maximize $I$. Figure \ref{fig:3}(e) shows that $I>90\%$ requires a cavity linewidth less than 124$\,$MHz. At a reflectivity of $99.95\%$\cite{10.1063/1.5026779}, this linewidth limits the free spectral range (FSR) to 779$\,$GHz. Compared with the free-space emission linewidth (5.41$\,$THz) this means that the spectral profile of the cavity would be comb-shaped, with the cavity sampling the emitter spectrum at multiples of the FSR. Single-photons originating from different comb peaks are of course distinguishable, so a high indistinguishability requires filtering out only one peak (for example with another cavity). This, however, is balanced by a loss in efficiency. To overcome this, the natural linewidth of the emitter into free-space must be narrowed. Cryogenic cooling is one option to narrow the linewidth sufficiently\cite{PhysRevB.98.081414}.

\section{Conclusion}
We have demonstrated coupling of a quantum emitter hosted by multilayer hBN to a confocal microcavity. The hemispherical geometries have been fabricated using FIB milling with sub-nm precision. The cavity mode volume is of the order of $\lambda^3$. The cavity improves the spectral purity of the emitter substantially, with the FWHM decreasing from 5.76 to 0.224$\,$nm. Moreover, the cavity suppresses off-resonant noise, which allows us to improve its single-photon purity. The excited state lifetime of the emitter is also shortened by the Purcell effect by a factor of 2.3. The emission of the cavity is linearly polarized and stable over long timeframes, with no signs of photobleaching or blinking. The cavity also features a linearly tunable PDMS spacer between both mirrors, which allows \textit{in-situ} tuning of the single-photon line over the full free-space ZPL of the quantum emitter. This would allow us to fabricate multiple identical single-photon sources, by locking all to the same emission wavelength, making this approach fully scalable. Furthermore, the complete SPS is portable and fully self-contained within $10\times 10\times 10\,$cm$^3$, the size of a 1U CubeSat. This makes the single-photon source a promising candidate for low cost satellite-based long-distance QKD, especially as the quantum emitters in hBN are space-certified. Despite the source's performance being not yet sufficient for one-way quantum computing, using the single-photon source for QKD even now enhances the quantum key generation rate on useful distances. The microcavity platform can also be easily adapted to other quantum emitters in 2D materials and offers a promising path towards scalable quantum information processing.

\begin{acknowledgments}
This work was funded by the Australian Research Council (CE110001027, FL150100019, DE140100805, DP180103238). We thank the ACT Node of the Australian National Fabrication Facility for access to their nano- and microfabrication facilities. We also thank Hark Hoe Tan for access to the TRPL system. R.L. acknowledges support by an Australian Government Research Training Program (RTP) Scholarship.
\end{acknowledgments}

\section*{Appendix: Methods}
\subsection{FIB milling}
Borosilicate glass substrates with a size of $18\times 18\,\text{mm}^2\times 160\,\mu\text{m}$ have been coated with 100$\,$nm gold using electron-beam thermal evaporation to prevent substrate charging effects. The ion accelerating voltage in the FIB (FEI Helios 600 NanoLab) is 30$\,$kV with currents $\leq 0.28\,$nA. The dose rate is encoded in the RGB color of a hemispherical pixel map. The dose rate to RGB value was carefully calibrated using AFM measurements. During the milling process we add I$_2$-gas, which ensures a smooth surface. Finally, the gold film is chemically etched using a custom-made potassium iodide (KI:I$_2$:H$_2$O with ratio 4:1:40 by weight) solution. Surface characterizations before and after the KI-etching show no difference in radius or roughness. We also tried hydrofluoric acid to etch the hemispheres, but for the feature sizes required for the cavity we could not achieve a smooth surface.
\subsection{Plasma sputtering}
We calibrated the deposition rate of the sputter coater (AJA) using variable angle spectroscopic ellipsometry (JA Woollam M-2000D), which measures film thickness and refractive index. At 565$\,$nm we found n$_\text{SiO2}=1.521$, n$_\text{TiO2}=2.135$ and n$_\text{MgF2}=1.390$. The deposition was done at room temperature. For the highly reflective coating we deposit alternating layers of SiO$_2$/TiO$_2$ with thickness of $\lambda/4n$ and the SiO$_2$ terminating the mirror. Due to the refractive index of MgF$_2$ being roughly in the middle between glass and air, the backsides of the substrates are coated with one quarter wave layer of MgF$_2$, serving as an anti-reflective coating. To maximize the escape efficiency into one particular direction it is common to make one of the stacks thicker (e.g. 10:9), so the photons couple primarily into a single direction. For simplicity, we used 9:9 stacks, which thus introduces 50$\%$ loss.
\subsection{Quantum emitter fabrication}
The flat mirrors have been coated with 300$\,$nm 950 PMMA A4. Multilayer hBN flakes have been exfoliated from bulk (HQGraphene) and transferred onto the PMMA layer by dry contact. Oxygen plasma etching (500$\,$W for 2$\,$min generated from a microwave field at a gas flow of 300$\,$ccm$^3$/min) removes the PMMA around the flake as well as creates the quantum emitters. The PMMA below the flake is decomposed during the annealing, which also stabilizes the optical emission properties (more details have been published previously\cite{10.1021/acsphotonics.8b00127}).
\subsection{Optical characterization}
Each flake has been scanned using a custom-built confocal micro-photoluminescence setup with a resolution of $0.5\,\mu$m and a spectrum has been recorded at each scanning position. The excitation laser with a wavelength of 522$\,$nm is non-resonant with the optical transition energy of the defect. The laser light is blocked with a Semrock RazorEdge ultrasteep long-pass edge filter. With a laser pulse length of 300$\,$fs at a repetition rate of 20.8$\,$MHz, the setup also allows us to measure the excited state lifetime. The pulses are split into trigger and excitation pulses, and the photoluminescence is detected by a SPAD (Micro Photon Devices). The time correlation between trigger pulse and arrival time of the photoluminescence is given by a time-to-digital converter (PicoQuant PicoHarp 300). The photoluminescence is coupled via a grating to the SPAD, which makes the TRPL wavelength-sensitive. This allows us to measure the lifetime of the ZPL only. The second-order correlation function measurements have been performed using two SPADs in the exit ports of a beam splitter and under continuous excitation. We fit the function
\begin{align}
g^{\left(2\right)}\left(\tau\right)=1-Ae^{-\abs{\tau}/t_1}+Be^{-\abs{\tau}/t_2}
\end{align}
with the anti- and bunching amplitudes $A$ and $B$, and the decay times $t_1$ and $t_2$. The experimental data is normalized such that $g^{\left(2\right)}\left(\tau\rightarrow\infty\right)=1$. The background corrected $g^{(2)}_\text{c}$ is given by
\begin{align}
g^{(2)}_\text{c}=\frac{g^{(2)}-(1-\rho^2)}{\rho^2}
\end{align}
with $\rho=\text{SNR}/(\text{SNR}+1)$ where SNR is the signal-to-noise ratio. In addition to the long-pass filter, the photoluminescence for correlation function measurements is band-pass filtered around the ZPL. We utilize linear variable filters (Delta Optical Thin Film 3G LVLWP and 3G LVSWP) to tune center and bandwidth of the band-pass filtering system.


\begin{thebibliography}{56}%
\makeatletter
\providecommand \@ifxundefined [1]{%
 \@ifx{#1\undefined}
}%
\providecommand \@ifnum [1]{%
 \ifnum #1\expandafter \@firstoftwo
 \else \expandafter \@secondoftwo
 \fi
}%
\providecommand \@ifx [1]{%
 \ifx #1\expandafter \@firstoftwo
 \else \expandafter \@secondoftwo
 \fi
}%
\providecommand \natexlab [1]{#1}%
\providecommand \enquote  [1]{``#1''}%
\providecommand \bibnamefont  [1]{#1}%
\providecommand \bibfnamefont [1]{#1}%
\providecommand \citenamefont [1]{#1}%
\providecommand \href@noop [0]{\@secondoftwo}%
\providecommand \href [0]{\begingroup \@sanitize@url \@href}%
\providecommand \@href[1]{\@@startlink{#1}\@@href}%
\providecommand \@@href[1]{\endgroup#1\@@endlink}%
\providecommand \@sanitize@url [0]{\catcode `\\12\catcode `\$12\catcode
  `\&12\catcode `\#12\catcode `\^12\catcode `\_12\catcode `\%12\relax}%
\providecommand \@@startlink[1]{}%
\providecommand \@@endlink[0]{}%
\providecommand \url  [0]{\begingroup\@sanitize@url \@url }%
\providecommand \@url [1]{\endgroup\@href {#1}{\urlprefix }}%
\providecommand \urlprefix  [0]{URL }%
\providecommand \Eprint [0]{\href }%
\providecommand \doibase [0]{http://dx.doi.org/}%
\providecommand \selectlanguage [0]{\@gobble}%
\providecommand \bibinfo  [0]{\@secondoftwo}%
\providecommand \bibfield  [0]{\@secondoftwo}%
\providecommand \translation [1]{[#1]}%
\providecommand \BibitemOpen [0]{}%
\providecommand \bibitemStop [0]{}%
\providecommand \bibitemNoStop [0]{.\EOS\space}%
\providecommand \EOS [0]{\spacefactor3000\relax}%
\providecommand \BibitemShut  [1]{\csname bibitem#1\endcsname}%
\let\auto@bib@innerbib\@empty
\bibitem [{\citenamefont {O'Brien}\ \emph {et~al.}(2009)\citenamefont
  {O'Brien}, \citenamefont {Furusawa},\ and\ \citenamefont {Vu{\v
  c}kovi{\'c}}}]{10.1038/nphoton.2009.229}%
  \BibitemOpen
  \bibfield  {author} {\bibinfo {author} {\bibfnamefont {Jeremy~L.}\
  \bibnamefont {O'Brien}}, \bibinfo {author} {\bibfnamefont {Akira}\
  \bibnamefont {Furusawa}}, \ and\ \bibinfo {author} {\bibfnamefont {Jelena}\
  \bibnamefont {Vu{\v c}kovi{\'c}}},\ }\bibfield  {title} {\enquote {\bibinfo
  {title} {Photonic quantum technologies},}\ }\href {\doibase
  10.1038/nphoton.2009.229} {\bibfield  {journal} {\bibinfo  {journal} {Nat.
  Photon.}\ }\textbf {\bibinfo {volume} {3}},\ \bibinfo {pages} {687--695}
  (\bibinfo {year} {2009})}\BibitemShut {NoStop}%
\bibitem [{\citenamefont {Senellart}\ \emph {et~al.}(2017)\citenamefont
  {Senellart}, \citenamefont {Solomon},\ and\ \citenamefont
  {White}}]{10.1038/nnano.2017.218}%
  \BibitemOpen
  \bibfield  {author} {\bibinfo {author} {\bibfnamefont {Pascale}\ \bibnamefont
  {Senellart}}, \bibinfo {author} {\bibfnamefont {Glenn}\ \bibnamefont
  {Solomon}}, \ and\ \bibinfo {author} {\bibfnamefont {Andrew}\ \bibnamefont
  {White}},\ }\bibfield  {title} {\enquote {\bibinfo {title} {High-performance
  semiconductor quantum-dot single-photon sources},}\ }\href {\doibase
  10.1038/nnano.2017.218} {\bibfield  {journal} {\bibinfo  {journal} {Nat.
  Nanotechnol.}\ }\textbf {\bibinfo {volume} {12}},\ \bibinfo {pages}
  {1026--1039} (\bibinfo {year} {2017})}\BibitemShut {NoStop}%
\bibitem [{\citenamefont {Barros}\ \emph {et~al.}(2009)\citenamefont {Barros},
  \citenamefont {Stute}, \citenamefont {Northup}, \citenamefont {Russo},
  \citenamefont {Schmidt},\ and\ \citenamefont
  {Blatt}}]{10.1088/1367-2630/11/10/103004}%
  \BibitemOpen
  \bibfield  {author} {\bibinfo {author} {\bibfnamefont {H.~G.}\ \bibnamefont
  {Barros}}, \bibinfo {author} {\bibfnamefont {A.}~\bibnamefont {Stute}},
  \bibinfo {author} {\bibfnamefont {T.~E.}\ \bibnamefont {Northup}}, \bibinfo
  {author} {\bibfnamefont {C.}~\bibnamefont {Russo}}, \bibinfo {author}
  {\bibfnamefont {P.~O.}\ \bibnamefont {Schmidt}}, \ and\ \bibinfo {author}
  {\bibfnamefont {R.}~\bibnamefont {Blatt}},\ }\bibfield  {title} {\enquote
  {\bibinfo {title} {Deterministic single-photon source from a single ion},}\
  }\href {\doibase 10.1088/1367-2630/11/10/103004} {\bibfield  {journal}
  {\bibinfo  {journal} {New J. of Phys.}\ }\textbf {\bibinfo {volume} {11}},\
  \bibinfo {pages} {103004} (\bibinfo {year} {2009})}\BibitemShut {NoStop}%
\bibitem [{\citenamefont {Aharonovich}\ \emph {et~al.}(2016)\citenamefont
  {Aharonovich}, \citenamefont {Englund},\ and\ \citenamefont
  {Toth}}]{10.1038/nphoton.2016.186}%
  \BibitemOpen
  \bibfield  {author} {\bibinfo {author} {\bibfnamefont {Igor}\ \bibnamefont
  {Aharonovich}}, \bibinfo {author} {\bibfnamefont {Dirk}\ \bibnamefont
  {Englund}}, \ and\ \bibinfo {author} {\bibfnamefont {Milos}\ \bibnamefont
  {Toth}},\ }\bibfield  {title} {\enquote {\bibinfo {title} {Solid-state
  single-photon emitters},}\ }\href {\doibase 10.1038/nphoton.2016.186}
  {\bibfield  {journal} {\bibinfo  {journal} {Nat. Photon.}\ }\textbf {\bibinfo
  {volume} {10}},\ \bibinfo {pages} {631--641} (\bibinfo {year}
  {2016})}\BibitemShut {NoStop}%
\bibitem [{\citenamefont {Bock}\ \emph {et~al.}(2016)\citenamefont {Bock},
  \citenamefont {Lenhard}, \citenamefont {Chunnilall},\ and\ \citenamefont
  {Becher}}]{10.1364/OE.24.023992}%
  \BibitemOpen
  \bibfield  {author} {\bibinfo {author} {\bibfnamefont {Matthias}\
  \bibnamefont {Bock}}, \bibinfo {author} {\bibfnamefont {Andreas}\
  \bibnamefont {Lenhard}}, \bibinfo {author} {\bibfnamefont {Christopher}\
  \bibnamefont {Chunnilall}}, \ and\ \bibinfo {author} {\bibfnamefont
  {Christoph}\ \bibnamefont {Becher}},\ }\bibfield  {title} {\enquote {\bibinfo
  {title} {Highly efficient heralded single-photon source for telecom
  wavelengths based on a {PPLN} waveguide},}\ }\href {\doibase
  10.1364/OE.24.023992} {\bibfield  {journal} {\bibinfo  {journal} {Opt.
  Express}\ }\textbf {\bibinfo {volume} {24}},\ \bibinfo {pages} {23992--24001}
  (\bibinfo {year} {2016})}\BibitemShut {NoStop}%
\bibitem [{\citenamefont {Tonndorf}\ \emph {et~al.}(2015)\citenamefont
  {Tonndorf}, \citenamefont {Schmidt}, \citenamefont {Schneider}, \citenamefont
  {Kern}, \citenamefont {Buscema}, \citenamefont {Steele}, \citenamefont
  {Castellanos-Gomez}, \citenamefont {van~der Zant}, \citenamefont
  {de~Vasconcellos},\ and\ \citenamefont
  {Bratschitsch}}]{10.1364/OPTICA.2.000347}%
  \BibitemOpen
  \bibfield  {author} {\bibinfo {author} {\bibfnamefont {Philipp}\ \bibnamefont
  {Tonndorf}}, \bibinfo {author} {\bibfnamefont {Robert}\ \bibnamefont
  {Schmidt}}, \bibinfo {author} {\bibfnamefont {Robert}\ \bibnamefont
  {Schneider}}, \bibinfo {author} {\bibfnamefont {Johannes}\ \bibnamefont
  {Kern}}, \bibinfo {author} {\bibfnamefont {Michele}\ \bibnamefont {Buscema}},
  \bibinfo {author} {\bibfnamefont {Gary~A.}\ \bibnamefont {Steele}}, \bibinfo
  {author} {\bibfnamefont {Andres}\ \bibnamefont {Castellanos-Gomez}}, \bibinfo
  {author} {\bibfnamefont {Herre S.~J.}\ \bibnamefont {van~der Zant}}, \bibinfo
  {author} {\bibfnamefont {Steffen~Michaelis}\ \bibnamefont {de~Vasconcellos}},
  \ and\ \bibinfo {author} {\bibfnamefont {Rudolf}\ \bibnamefont
  {Bratschitsch}},\ }\bibfield  {title} {\enquote {\bibinfo {title}
  {Single-photon emission from localized excitons in an atomically thin
  semiconductor},}\ }\href {\doibase 10.1364/OPTICA.2.000347} {\bibfield
  {journal} {\bibinfo  {journal} {Optica}\ }\textbf {\bibinfo {volume} {2}},\
  \bibinfo {pages} {347--352} (\bibinfo {year} {2015})}\BibitemShut {NoStop}%
\bibitem [{\citenamefont {Srivastava}\ \emph {et~al.}(2015)\citenamefont
  {Srivastava}, \citenamefont {Sidler}, \citenamefont {Allain}, \citenamefont
  {Lembke}, \citenamefont {Kis},\ and\ \citenamefont
  {Imamo\v{g}lu}}]{nnano.2015.60}%
  \BibitemOpen
  \bibfield  {author} {\bibinfo {author} {\bibfnamefont {A.}~\bibnamefont
  {Srivastava}}, \bibinfo {author} {\bibfnamefont {M.}~\bibnamefont {Sidler}},
  \bibinfo {author} {\bibfnamefont {A.~V.}\ \bibnamefont {Allain}}, \bibinfo
  {author} {\bibfnamefont {D.~S.}\ \bibnamefont {Lembke}}, \bibinfo {author}
  {\bibfnamefont {A.}~\bibnamefont {Kis}}, \ and\ \bibinfo {author}
  {\bibfnamefont {A.}~\bibnamefont {Imamo\v{g}lu}},\ }\bibfield  {title}
  {\enquote {\bibinfo {title} {Optically active quantum dots in monolayer
  {WSe2}},}\ }\href {\doibase 10.1038/nnano.2015.60} {\bibfield  {journal}
  {\bibinfo  {journal} {Nat. Nanotechnol.}\ }\textbf {\bibinfo {volume} {10}},\
  \bibinfo {pages} {491--496} (\bibinfo {year} {2015})}\BibitemShut {NoStop}%
\bibitem [{\citenamefont {Koperski}\ \emph {et~al.}(2015)\citenamefont
  {Koperski}, \citenamefont {Nogajewski}, \citenamefont {Arora}, \citenamefont
  {Cherkez}, \citenamefont {Mallet}, \citenamefont {Veuillen}, \citenamefont
  {Marcus}, \citenamefont {Kossacki},\ and\ \citenamefont
  {Potemski}}]{nnano.2015.67}%
  \BibitemOpen
  \bibfield  {author} {\bibinfo {author} {\bibfnamefont {M.}~\bibnamefont
  {Koperski}}, \bibinfo {author} {\bibfnamefont {K.}~\bibnamefont
  {Nogajewski}}, \bibinfo {author} {\bibfnamefont {A.}~\bibnamefont {Arora}},
  \bibinfo {author} {\bibfnamefont {V.}~\bibnamefont {Cherkez}}, \bibinfo
  {author} {\bibfnamefont {P.}~\bibnamefont {Mallet}}, \bibinfo {author}
  {\bibfnamefont {J.-Y.}\ \bibnamefont {Veuillen}}, \bibinfo {author}
  {\bibfnamefont {J.}~\bibnamefont {Marcus}}, \bibinfo {author} {\bibfnamefont
  {P.}~\bibnamefont {Kossacki}}, \ and\ \bibinfo {author} {\bibfnamefont
  {M.}~\bibnamefont {Potemski}},\ }\bibfield  {title} {\enquote {\bibinfo
  {title} {Single photon emitters in exfoliated {WSe2} structures},}\ }\href
  {\doibase 10.1038/nnano.2015.67} {\bibfield  {journal} {\bibinfo  {journal}
  {Nat. Nanotechnol.}\ }\textbf {\bibinfo {volume} {10}},\ \bibinfo {pages}
  {503--506} (\bibinfo {year} {2015})}\BibitemShut {NoStop}%
\bibitem [{\citenamefont {He}\ \emph {et~al.}(2015)\citenamefont {He},
  \citenamefont {Clark}, \citenamefont {Schaibley}, \citenamefont {He},
  \citenamefont {Chen}, \citenamefont {Wei}, \citenamefont {Ding},
  \citenamefont {Zhang}, \citenamefont {Yao}, \citenamefont {Xu}, \citenamefont
  {Lu},\ and\ \citenamefont {Pan}}]{nnano.2015.75}%
  \BibitemOpen
  \bibfield  {author} {\bibinfo {author} {\bibfnamefont {Y.-M.}\ \bibnamefont
  {He}}, \bibinfo {author} {\bibfnamefont {G.}~\bibnamefont {Clark}}, \bibinfo
  {author} {\bibfnamefont {J.~R.}\ \bibnamefont {Schaibley}}, \bibinfo {author}
  {\bibfnamefont {Y.}~\bibnamefont {He}}, \bibinfo {author} {\bibfnamefont
  {M.-C.}\ \bibnamefont {Chen}}, \bibinfo {author} {\bibfnamefont {Y.-J.}\
  \bibnamefont {Wei}}, \bibinfo {author} {\bibfnamefont {X.}~\bibnamefont
  {Ding}}, \bibinfo {author} {\bibfnamefont {Q.}~\bibnamefont {Zhang}},
  \bibinfo {author} {\bibfnamefont {W.}~\bibnamefont {Yao}}, \bibinfo {author}
  {\bibfnamefont {X.}~\bibnamefont {Xu}}, \bibinfo {author} {\bibfnamefont
  {C.-Y.}\ \bibnamefont {Lu}}, \ and\ \bibinfo {author} {\bibfnamefont {J.-W.}\
  \bibnamefont {Pan}},\ }\bibfield  {title} {\enquote {\bibinfo {title} {Single
  quantum emitters in monolayer semiconductors},}\ }\href {\doibase
  10.1038/nnano.2015.75} {\bibfield  {journal} {\bibinfo  {journal} {Nat.
  Nanotechnol.}\ }\textbf {\bibinfo {volume} {10}},\ \bibinfo {pages}
  {497--502} (\bibinfo {year} {2015})}\BibitemShut {NoStop}%
\bibitem [{\citenamefont {Chakraborty}\ \emph {et~al.}(2015)\citenamefont
  {Chakraborty}, \citenamefont {Kinnischtzke}, \citenamefont {Goodfellow},
  \citenamefont {Beams},\ and\ \citenamefont {Vamivakas}}]{nnano.2015.79}%
  \BibitemOpen
  \bibfield  {author} {\bibinfo {author} {\bibfnamefont {C.}~\bibnamefont
  {Chakraborty}}, \bibinfo {author} {\bibfnamefont {L.}~\bibnamefont
  {Kinnischtzke}}, \bibinfo {author} {\bibfnamefont {K.~M.}\ \bibnamefont
  {Goodfellow}}, \bibinfo {author} {\bibfnamefont {R.}~\bibnamefont {Beams}}, \
  and\ \bibinfo {author} {\bibfnamefont {A.~N.}\ \bibnamefont {Vamivakas}},\
  }\bibfield  {title} {\enquote {\bibinfo {title} {Voltage-controlled quantum
  light from an atomically thin semiconductor},}\ }\href {\doibase
  10.1038/nnano.2015.79} {\bibfield  {journal} {\bibinfo  {journal} {Nat.
  Nanotechnol.}\ }\textbf {\bibinfo {volume} {10}},\ \bibinfo {pages}
  {507--511} (\bibinfo {year} {2015})}\BibitemShut {NoStop}%
\bibitem [{\citenamefont {Palacios-Berraquero}\ \emph
  {et~al.}(2016)\citenamefont {Palacios-Berraquero}, \citenamefont {Barbone},
  \citenamefont {Kara}, \citenamefont {Chen}, \citenamefont {Goykhman},
  \citenamefont {Yoon}, \citenamefont {Ott}, \citenamefont {Beitner},
  \citenamefont {Watanabe}, \citenamefont {Taniguchi}, \citenamefont
  {Ferrari},\ and\ \citenamefont {Atat{\"u}re}}]{10.1038/ncomms12978}%
  \BibitemOpen
  \bibfield  {author} {\bibinfo {author} {\bibfnamefont {Carmen}\ \bibnamefont
  {Palacios-Berraquero}}, \bibinfo {author} {\bibfnamefont {Matteo}\
  \bibnamefont {Barbone}}, \bibinfo {author} {\bibfnamefont {Dhiren~M.}\
  \bibnamefont {Kara}}, \bibinfo {author} {\bibfnamefont {Xiaolong}\
  \bibnamefont {Chen}}, \bibinfo {author} {\bibfnamefont {Ilya}\ \bibnamefont
  {Goykhman}}, \bibinfo {author} {\bibfnamefont {Duhee}\ \bibnamefont {Yoon}},
  \bibinfo {author} {\bibfnamefont {Anna~K.}\ \bibnamefont {Ott}}, \bibinfo
  {author} {\bibfnamefont {Jan}\ \bibnamefont {Beitner}}, \bibinfo {author}
  {\bibfnamefont {Kenji}\ \bibnamefont {Watanabe}}, \bibinfo {author}
  {\bibfnamefont {Takashi}\ \bibnamefont {Taniguchi}}, \bibinfo {author}
  {\bibfnamefont {Andrea~C.}\ \bibnamefont {Ferrari}}, \ and\ \bibinfo {author}
  {\bibfnamefont {Mete}\ \bibnamefont {Atat{\"u}re}},\ }\bibfield  {title}
  {\enquote {\bibinfo {title} {Atomically thin quantum light-emitting
  diodes},}\ }\href {\doibase 10.1038/ncomms12978} {\bibfield  {journal}
  {\bibinfo  {journal} {Nat. Commun.}\ }\textbf {\bibinfo {volume} {7}},\
  \bibinfo {pages} {12978} (\bibinfo {year} {2016})}\BibitemShut {NoStop}%
\bibitem [{\citenamefont {Branny}\ \emph {et~al.}(2016)\citenamefont {Branny},
  \citenamefont {Wang}, \citenamefont {Kumar}, \citenamefont {Robert},
  \citenamefont {Lassagne}, \citenamefont {Marie}, \citenamefont {Gerardot},\
  and\ \citenamefont {Urbaszek}}]{10.1063/1.4945268}%
  \BibitemOpen
  \bibfield  {author} {\bibinfo {author} {\bibfnamefont {Artur}\ \bibnamefont
  {Branny}}, \bibinfo {author} {\bibfnamefont {Gang}\ \bibnamefont {Wang}},
  \bibinfo {author} {\bibfnamefont {Santosh}\ \bibnamefont {Kumar}}, \bibinfo
  {author} {\bibfnamefont {Cedric}\ \bibnamefont {Robert}}, \bibinfo {author}
  {\bibfnamefont {Benjamin}\ \bibnamefont {Lassagne}}, \bibinfo {author}
  {\bibfnamefont {Xavier}\ \bibnamefont {Marie}}, \bibinfo {author}
  {\bibfnamefont {Brian~D.}\ \bibnamefont {Gerardot}}, \ and\ \bibinfo {author}
  {\bibfnamefont {Bernhard}\ \bibnamefont {Urbaszek}},\ }\bibfield  {title}
  {\enquote {\bibinfo {title} {Discrete quantum dot like emitters in monolayer
  {MoSe2}: Spatial mapping, magneto-optics, and charge tuning},}\ }\href
  {\doibase 10.1063/1.4945268} {\bibfield  {journal} {\bibinfo  {journal}
  {Appl. Phys. Lett.}\ }\textbf {\bibinfo {volume} {108}},\ \bibinfo {pages}
  {142101} (\bibinfo {year} {2016})}\BibitemShut {NoStop}%
\bibitem [{\citenamefont {Klein}\ \emph {et~al.}(2019)\citenamefont {Klein},
  \citenamefont {Lorke}, \citenamefont {Florian}, \citenamefont {Sigger},
  \citenamefont {Wierzbowski}, \citenamefont {Cerne}, \citenamefont
  {M{\"u}ller}, \citenamefont {Taniguchi}, \citenamefont {Watanabe},
  \citenamefont {Wurstbauer}, \citenamefont {Kaniber}, \citenamefont {Knap},
  \citenamefont {Schmidt}, \citenamefont {Finley},\ and\ \citenamefont
  {Holleitner}}]{arXiv:1901.01042}%
  \BibitemOpen
  \bibfield  {author} {\bibinfo {author} {\bibfnamefont {J.}~\bibnamefont
  {Klein}}, \bibinfo {author} {\bibfnamefont {M.}~\bibnamefont {Lorke}},
  \bibinfo {author} {\bibfnamefont {M.}~\bibnamefont {Florian}}, \bibinfo
  {author} {\bibfnamefont {F.}~\bibnamefont {Sigger}}, \bibinfo {author}
  {\bibfnamefont {J.}~\bibnamefont {Wierzbowski}}, \bibinfo {author}
  {\bibfnamefont {J.}~\bibnamefont {Cerne}}, \bibinfo {author} {\bibfnamefont
  {K.}~\bibnamefont {M{\"u}ller}}, \bibinfo {author} {\bibfnamefont
  {T.}~\bibnamefont {Taniguchi}}, \bibinfo {author} {\bibfnamefont
  {K.}~\bibnamefont {Watanabe}}, \bibinfo {author} {\bibfnamefont
  {U.}~\bibnamefont {Wurstbauer}}, \bibinfo {author} {\bibfnamefont
  {M.}~\bibnamefont {Kaniber}}, \bibinfo {author} {\bibfnamefont
  {M.}~\bibnamefont {Knap}}, \bibinfo {author} {\bibfnamefont {R.}~\bibnamefont
  {Schmidt}}, \bibinfo {author} {\bibfnamefont {J.~J.}\ \bibnamefont {Finley}},
  \ and\ \bibinfo {author} {\bibfnamefont {A.~W.}\ \bibnamefont {Holleitner}},\
  }\bibfield  {title} {\enquote {\bibinfo {title} {Atomistic defect states as
  quantum emitters in monolayer {MoS$_2$}},}\ }\href
  {https://arxiv.org/abs/1901.01042} {\  (\bibinfo {year} {2019})},\ \bibinfo
  {note} {arXiv:1901.01042},\ \Eprint {http://arxiv.org/abs/arXiv:1901.01042}
  {arXiv:1901.01042} \BibitemShut {NoStop}%
\bibitem [{\citenamefont {Tran}\ \emph
  {et~al.}(2016{\natexlab{a}})\citenamefont {Tran}, \citenamefont {Bray},
  \citenamefont {Ford}, \citenamefont {Toth},\ and\ \citenamefont
  {Aharonovich}}]{nnano.2015.242}%
  \BibitemOpen
  \bibfield  {author} {\bibinfo {author} {\bibfnamefont {T.~T.}\ \bibnamefont
  {Tran}}, \bibinfo {author} {\bibfnamefont {K.}~\bibnamefont {Bray}}, \bibinfo
  {author} {\bibfnamefont {M.~J.}\ \bibnamefont {Ford}}, \bibinfo {author}
  {\bibfnamefont {M.}~\bibnamefont {Toth}}, \ and\ \bibinfo {author}
  {\bibfnamefont {I.}~\bibnamefont {Aharonovich}},\ }\bibfield  {title}
  {\enquote {\bibinfo {title} {Quantum emission from hexagonal boron nitride
  monolayers},}\ }\href {\doibase 10.1038/nnano.2015.242} {\bibfield  {journal}
  {\bibinfo  {journal} {Nat. Nanotechnol.}\ }\textbf {\bibinfo {volume} {11}},\
  \bibinfo {pages} {37--41} (\bibinfo {year} {2016}{\natexlab{a}})}\BibitemShut
  {NoStop}%
\bibitem [{\citenamefont {Tran}\ \emph
  {et~al.}(2016{\natexlab{b}})\citenamefont {Tran}, \citenamefont {Elbadawi},
  \citenamefont {Totonjian}, \citenamefont {Lobo}, \citenamefont {Grosso},
  \citenamefont {Moon}, \citenamefont {Englund}, \citenamefont {Ford},
  \citenamefont {Aharonovich},\ and\ \citenamefont
  {Toth}}]{10.1021/acsnano.6b03602}%
  \BibitemOpen
  \bibfield  {author} {\bibinfo {author} {\bibfnamefont {Toan~Trong}\
  \bibnamefont {Tran}}, \bibinfo {author} {\bibfnamefont {Christopher}\
  \bibnamefont {Elbadawi}}, \bibinfo {author} {\bibfnamefont {Daniel}\
  \bibnamefont {Totonjian}}, \bibinfo {author} {\bibfnamefont {Charlene~J.}\
  \bibnamefont {Lobo}}, \bibinfo {author} {\bibfnamefont {Gabriele}\
  \bibnamefont {Grosso}}, \bibinfo {author} {\bibfnamefont {Hyowon}\
  \bibnamefont {Moon}}, \bibinfo {author} {\bibfnamefont {Dirk~R.}\
  \bibnamefont {Englund}}, \bibinfo {author} {\bibfnamefont {Michael~J.}\
  \bibnamefont {Ford}}, \bibinfo {author} {\bibfnamefont {Igor}\ \bibnamefont
  {Aharonovich}}, \ and\ \bibinfo {author} {\bibfnamefont {Milos}\ \bibnamefont
  {Toth}},\ }\bibfield  {title} {\enquote {\bibinfo {title} {Robust multicolor
  single photon emission from point defects in hexagonal boron nitride},}\
  }\href {\doibase 10.1021/acsnano.6b03602} {\bibfield  {journal} {\bibinfo
  {journal} {ACS Nano}\ }\textbf {\bibinfo {volume} {10}},\ \bibinfo {pages}
  {7331--7338} (\bibinfo {year} {2016}{\natexlab{b}})}\BibitemShut {NoStop}%
\bibitem [{\citenamefont {Tran}\ \emph
  {et~al.}(2016{\natexlab{c}})\citenamefont {Tran}, \citenamefont {Zachreson},
  \citenamefont {Berhane}, \citenamefont {Bray}, \citenamefont {Sandstrom},
  \citenamefont {Li}, \citenamefont {Taniguchi}, \citenamefont {Watanabe},
  \citenamefont {Aharonovich},\ and\ \citenamefont
  {Toth}}]{10.1103/PhysRevApplied.5.034005}%
  \BibitemOpen
  \bibfield  {author} {\bibinfo {author} {\bibfnamefont {Toan~Trong}\
  \bibnamefont {Tran}}, \bibinfo {author} {\bibfnamefont {Cameron}\
  \bibnamefont {Zachreson}}, \bibinfo {author} {\bibfnamefont
  {Amanuel~Michael}\ \bibnamefont {Berhane}}, \bibinfo {author} {\bibfnamefont
  {Kerem}\ \bibnamefont {Bray}}, \bibinfo {author} {\bibfnamefont
  {Russell~Guy}\ \bibnamefont {Sandstrom}}, \bibinfo {author} {\bibfnamefont
  {Lu~Hua}\ \bibnamefont {Li}}, \bibinfo {author} {\bibfnamefont {Takashi}\
  \bibnamefont {Taniguchi}}, \bibinfo {author} {\bibfnamefont {Kenji}\
  \bibnamefont {Watanabe}}, \bibinfo {author} {\bibfnamefont {Igor}\
  \bibnamefont {Aharonovich}}, \ and\ \bibinfo {author} {\bibfnamefont {Milos}\
  \bibnamefont {Toth}},\ }\bibfield  {title} {\enquote {\bibinfo {title}
  {Quantum emission from defects in single-crystalline hexagonal boron
  nitride},}\ }\href {\doibase 10.1103/PhysRevApplied.5.034005} {\bibfield
  {journal} {\bibinfo  {journal} {Phys. Rev. Applied}\ }\textbf {\bibinfo
  {volume} {5}},\ \bibinfo {pages} {034005} (\bibinfo {year}
  {2016}{\natexlab{c}})}\BibitemShut {NoStop}%
\bibitem [{\citenamefont {Cassabois}\ \emph {et~al.}(2016)\citenamefont
  {Cassabois}, \citenamefont {Valvin},\ and\ \citenamefont
  {B.Gil}}]{10.1038/nphoton2015.77}%
  \BibitemOpen
  \bibfield  {author} {\bibinfo {author} {\bibfnamefont {G.}~\bibnamefont
  {Cassabois}}, \bibinfo {author} {\bibfnamefont {P.}~\bibnamefont {Valvin}}, \
  and\ \bibinfo {author} {\bibnamefont {B.Gil}},\ }\bibfield  {title} {\enquote
  {\bibinfo {title} {Hexagonal boron nitride is an indirect bandgap
  semiconductor},}\ }\href {\doibase 10.1038/nphoton2015.77} {\bibfield
  {journal} {\bibinfo  {journal} {Nat. Photonics}\ }\textbf {\bibinfo {volume}
  {10}},\ \bibinfo {pages} {262--266} (\bibinfo {year} {2016})}\BibitemShut
  {NoStop}%
\bibitem [{\citenamefont {Schell}\ \emph {et~al.}(2017)\citenamefont {Schell},
  \citenamefont {Takashima}, \citenamefont {Tran}, \citenamefont
  {Aharonovich},\ and\ \citenamefont
  {Takeuchi}}]{10.1021/acsphotonics.7b00025}%
  \BibitemOpen
  \bibfield  {author} {\bibinfo {author} {\bibfnamefont {Andreas~W.}\
  \bibnamefont {Schell}}, \bibinfo {author} {\bibfnamefont {Hideaki}\
  \bibnamefont {Takashima}}, \bibinfo {author} {\bibfnamefont {Toan~Trong}\
  \bibnamefont {Tran}}, \bibinfo {author} {\bibfnamefont {Igor}\ \bibnamefont
  {Aharonovich}}, \ and\ \bibinfo {author} {\bibfnamefont {Shigeki}\
  \bibnamefont {Takeuchi}},\ }\bibfield  {title} {\enquote {\bibinfo {title}
  {Coupling quantum emitters in {2D} materials with tapered fibers},}\ }\href
  {\doibase 10.1021/acsphotonics.7b00025} {\bibfield  {journal} {\bibinfo
  {journal} {ACS Photonics}\ }\textbf {\bibinfo {volume} {4}},\ \bibinfo
  {pages} {761--767} (\bibinfo {year} {2017})}\BibitemShut {NoStop}%
\bibitem [{\citenamefont {Vogl}\ \emph {et~al.}(2017)\citenamefont {Vogl},
  \citenamefont {Lu},\ and\ \citenamefont {Lam}}]{10.1088/1361-6463/aa7839}%
  \BibitemOpen
  \bibfield  {author} {\bibinfo {author} {\bibfnamefont {Tobias}\ \bibnamefont
  {Vogl}}, \bibinfo {author} {\bibfnamefont {Yuerui}\ \bibnamefont {Lu}}, \
  and\ \bibinfo {author} {\bibfnamefont {Ping~Koy}\ \bibnamefont {Lam}},\
  }\bibfield  {title} {\enquote {\bibinfo {title} {Room temperature single
  photon source using fiber-integrated hexagonal boron nitride},}\ }\href
  {https://doi.org/10.1088/1361-6463/aa7839} {\bibfield  {journal} {\bibinfo
  {journal} {J. Phys. D.}\ }\textbf {\bibinfo {volume} {50}},\ \bibinfo {pages}
  {295101} (\bibinfo {year} {2017})}\BibitemShut {NoStop}%
\bibitem [{\citenamefont {Vogl}\ \emph
  {et~al.}(2018{\natexlab{a}})\citenamefont {Vogl}, \citenamefont {Campbell},
  \citenamefont {Buchler}, \citenamefont {Lu},\ and\ \citenamefont
  {Lam}}]{10.1021/acsphotonics.8b00127}%
  \BibitemOpen
  \bibfield  {author} {\bibinfo {author} {\bibfnamefont {Tobias}\ \bibnamefont
  {Vogl}}, \bibinfo {author} {\bibfnamefont {Geoff}\ \bibnamefont {Campbell}},
  \bibinfo {author} {\bibfnamefont {Ben~C.}\ \bibnamefont {Buchler}}, \bibinfo
  {author} {\bibfnamefont {Yuerui}\ \bibnamefont {Lu}}, \ and\ \bibinfo
  {author} {\bibfnamefont {Ping~Koy}\ \bibnamefont {Lam}},\ }\bibfield  {title}
  {\enquote {\bibinfo {title} {Fabrication and deterministic transfer of
  high-quality quantum emitters in hexagonal boron nitride},}\ }\href {\doibase
  10.1021/acsphotonics.8b00127} {\bibfield  {journal} {\bibinfo  {journal} {ACS
  Photonics}\ }\textbf {\bibinfo {volume} {5}},\ \bibinfo {pages} {2305--2312}
  (\bibinfo {year} {2018}{\natexlab{a}})}\BibitemShut {NoStop}%
\bibitem [{\citenamefont {Kianinia}\ \emph {et~al.}(2017)\citenamefont
  {Kianinia}, \citenamefont {Regan}, \citenamefont {Tawfik}, \citenamefont
  {Tran}, \citenamefont {Ford}, \citenamefont {Aharonovich},\ and\
  \citenamefont {Toth}}]{10.1021/acsphotonics.7b00086}%
  \BibitemOpen
  \bibfield  {author} {\bibinfo {author} {\bibfnamefont {Mehran}\ \bibnamefont
  {Kianinia}}, \bibinfo {author} {\bibfnamefont {Blake}\ \bibnamefont {Regan}},
  \bibinfo {author} {\bibfnamefont {Sherif~Abdulkader}\ \bibnamefont {Tawfik}},
  \bibinfo {author} {\bibfnamefont {Toan~Trong}\ \bibnamefont {Tran}}, \bibinfo
  {author} {\bibfnamefont {Michael~J.}\ \bibnamefont {Ford}}, \bibinfo {author}
  {\bibfnamefont {Igor}\ \bibnamefont {Aharonovich}}, \ and\ \bibinfo {author}
  {\bibfnamefont {Milos}\ \bibnamefont {Toth}},\ }\bibfield  {title} {\enquote
  {\bibinfo {title} {Robust solid-state quantum system operating at 800 {K}},}\
  }\href {\doibase 10.1021/acsphotonics.7b00086} {\bibfield  {journal}
  {\bibinfo  {journal} {ACS Photonics}\ }\textbf {\bibinfo {volume} {4}},\
  \bibinfo {pages} {768--773} (\bibinfo {year} {2017})}\BibitemShut {NoStop}%
\bibitem [{\citenamefont {Vogl}\ \emph
  {et~al.}(2018{\natexlab{b}})\citenamefont {Vogl}, \citenamefont {Sripathy},
  \citenamefont {Sharma}, \citenamefont {Reddy}, \citenamefont {Sullivan},
  \citenamefont {Machacek}, \citenamefont {Zhang}, \citenamefont {Karouta},
  \citenamefont {Buchler}, \citenamefont {Doherty}, \citenamefont {Lu},\ and\
  \citenamefont {Lam}}]{arXiv:1811.10138}%
  \BibitemOpen
  \bibfield  {author} {\bibinfo {author} {\bibfnamefont {T.}~\bibnamefont
  {Vogl}}, \bibinfo {author} {\bibfnamefont {K.}~\bibnamefont {Sripathy}},
  \bibinfo {author} {\bibfnamefont {A.}~\bibnamefont {Sharma}}, \bibinfo
  {author} {\bibfnamefont {P.}~\bibnamefont {Reddy}}, \bibinfo {author}
  {\bibfnamefont {J.}~\bibnamefont {Sullivan}}, \bibinfo {author}
  {\bibfnamefont {J.~R.}\ \bibnamefont {Machacek}}, \bibinfo {author}
  {\bibfnamefont {L.}~\bibnamefont {Zhang}}, \bibinfo {author} {\bibfnamefont
  {F.}~\bibnamefont {Karouta}}, \bibinfo {author} {\bibfnamefont {B.~C.}\
  \bibnamefont {Buchler}}, \bibinfo {author} {\bibfnamefont {M.~W.}\
  \bibnamefont {Doherty}}, \bibinfo {author} {\bibfnamefont {Y.}~\bibnamefont
  {Lu}}, \ and\ \bibinfo {author} {\bibfnamefont {P.~K.}\ \bibnamefont {Lam}},\
  }\bibfield  {title} {\enquote {\bibinfo {title} {Radiation tolerance of
  two-dimensional material-based devices for space applications},}\ }\href
  {https://arxiv.org/abs/1811.10138} {\  (\bibinfo {year}
  {2018}{\natexlab{b}})},\ \bibinfo {note} {arXiv:1811.10138},\ \Eprint
  {http://arxiv.org/abs/arXiv:1811.10138} {arXiv:1811.10138} \BibitemShut
  {NoStop}%
\bibitem [{\citenamefont {Tawfik}\ \emph {et~al.}(2017)\citenamefont {Tawfik},
  \citenamefont {Ali}, \citenamefont {Fronzi}, \citenamefont {Kianinia},
  \citenamefont {Tran}, \citenamefont {Stampfl}, \citenamefont {Aharonovich},
  \citenamefont {Toth},\ and\ \citenamefont {Ford}}]{10.1039/C7NR04270A}%
  \BibitemOpen
  \bibfield  {author} {\bibinfo {author} {\bibfnamefont {Sherif~Abdulkader}\
  \bibnamefont {Tawfik}}, \bibinfo {author} {\bibfnamefont {Sajid}\
  \bibnamefont {Ali}}, \bibinfo {author} {\bibfnamefont {Marco}\ \bibnamefont
  {Fronzi}}, \bibinfo {author} {\bibfnamefont {Mehran}\ \bibnamefont
  {Kianinia}}, \bibinfo {author} {\bibfnamefont {Toan~Trong}\ \bibnamefont
  {Tran}}, \bibinfo {author} {\bibfnamefont {Catherine}\ \bibnamefont
  {Stampfl}}, \bibinfo {author} {\bibfnamefont {Igor}\ \bibnamefont
  {Aharonovich}}, \bibinfo {author} {\bibfnamefont {Milos}\ \bibnamefont
  {Toth}}, \ and\ \bibinfo {author} {\bibfnamefont {Michael~J.}\ \bibnamefont
  {Ford}},\ }\bibfield  {title} {\enquote {\bibinfo {title} {First-principles
  investigation of quantum emission from hbn defects},}\ }\href {\doibase
  10.1039/C7NR04270A} {\bibfield  {journal} {\bibinfo  {journal} {Nanoscale}\
  }\textbf {\bibinfo {volume} {9}},\ \bibinfo {pages} {13575--13582} (\bibinfo
  {year} {2017})}\BibitemShut {NoStop}%
\bibitem [{\citenamefont {Abdi}\ \emph {et~al.}(2018)\citenamefont {Abdi},
  \citenamefont {Chou}, \citenamefont {Gali},\ and\ \citenamefont
  {Plenio}}]{10.1021/acsphotonics.7b01442}%
  \BibitemOpen
  \bibfield  {author} {\bibinfo {author} {\bibfnamefont {Mehdi}\ \bibnamefont
  {Abdi}}, \bibinfo {author} {\bibfnamefont {Jyh-Pin}\ \bibnamefont {Chou}},
  \bibinfo {author} {\bibfnamefont {Adam}\ \bibnamefont {Gali}}, \ and\
  \bibinfo {author} {\bibfnamefont {Martin~B.}\ \bibnamefont {Plenio}},\
  }\bibfield  {title} {\enquote {\bibinfo {title} {Color centers in hexagonal
  boron nitride monolayers: A group theory and ab initio analysis},}\ }\href
  {\doibase 10.1021/acsphotonics.7b01442} {\bibfield  {journal} {\bibinfo
  {journal} {ACS Photonics}\ }\textbf {\bibinfo {volume} {5}},\ \bibinfo
  {pages} {1967--1976} (\bibinfo {year} {2018})}\BibitemShut {NoStop}%
\bibitem [{\citenamefont {Bourrellier}\ \emph {et~al.}(2016)\citenamefont
  {Bourrellier}, \citenamefont {Meuret}, \citenamefont {Tararan}, \citenamefont
  {St{\'e}phan}, \citenamefont {Kociak}, \citenamefont {Tizei},\ and\
  \citenamefont {Zobelli}}]{10.1021/acs.nanolett.6b01368}%
  \BibitemOpen
  \bibfield  {author} {\bibinfo {author} {\bibfnamefont {Romain}\ \bibnamefont
  {Bourrellier}}, \bibinfo {author} {\bibfnamefont {Sophie}\ \bibnamefont
  {Meuret}}, \bibinfo {author} {\bibfnamefont {Anna}\ \bibnamefont {Tararan}},
  \bibinfo {author} {\bibfnamefont {Odile}\ \bibnamefont {St{\'e}phan}},
  \bibinfo {author} {\bibfnamefont {Mathieu}\ \bibnamefont {Kociak}}, \bibinfo
  {author} {\bibfnamefont {Luiz H.~G.}\ \bibnamefont {Tizei}}, \ and\ \bibinfo
  {author} {\bibfnamefont {Alberto}\ \bibnamefont {Zobelli}},\ }\bibfield
  {title} {\enquote {\bibinfo {title} {Bright {UV} single photon emission at
  point defects in {h-BN}},}\ }\href {\doibase 10.1021/acs.nanolett.6b01368}
  {\bibfield  {journal} {\bibinfo  {journal} {Nano Lett.}\ }\textbf {\bibinfo
  {volume} {16}},\ \bibinfo {pages} {4317--4321} (\bibinfo {year}
  {2016})}\BibitemShut {NoStop}%
\bibitem [{\citenamefont {Shotan}\ \emph {et~al.}(2016)\citenamefont {Shotan},
  \citenamefont {Jayakumar}, \citenamefont {Considine}, \citenamefont
  {Mackoit}, \citenamefont {Fedder}, \citenamefont {Wrachtrup}, \citenamefont
  {Alkauskas}, \citenamefont {Doherty}, \citenamefont {Menon},\ and\
  \citenamefont {Meriles}}]{10.1021/acsphotonics.6b00736}%
  \BibitemOpen
  \bibfield  {author} {\bibinfo {author} {\bibfnamefont {Zav}\ \bibnamefont
  {Shotan}}, \bibinfo {author} {\bibfnamefont {Harishankar}\ \bibnamefont
  {Jayakumar}}, \bibinfo {author} {\bibfnamefont {Christopher~R.}\ \bibnamefont
  {Considine}}, \bibinfo {author} {\bibfnamefont {Ma{\v z}ena}\ \bibnamefont
  {Mackoit}}, \bibinfo {author} {\bibfnamefont {Helmut}\ \bibnamefont
  {Fedder}}, \bibinfo {author} {\bibfnamefont {J{\" o}rg}\ \bibnamefont
  {Wrachtrup}}, \bibinfo {author} {\bibfnamefont {Audrius}\ \bibnamefont
  {Alkauskas}}, \bibinfo {author} {\bibfnamefont {Marcus~W.}\ \bibnamefont
  {Doherty}}, \bibinfo {author} {\bibfnamefont {Vinod~M.}\ \bibnamefont
  {Menon}}, \ and\ \bibinfo {author} {\bibfnamefont {Carlos~A.}\ \bibnamefont
  {Meriles}},\ }\bibfield  {title} {\enquote {\bibinfo {title} {Photoinduced
  modification of single-photon emitters in hexagonal boron nitride},}\ }\href
  {\doibase 10.1021/acsphotonics.6b00736} {\bibfield  {journal} {\bibinfo
  {journal} {ACS Photonics}\ }\textbf {\bibinfo {volume} {3}},\ \bibinfo
  {pages} {2490--2496} (\bibinfo {year} {2016})}\BibitemShut {NoStop}%
\bibitem [{\citenamefont {Chejanovsky}\ \emph {et~al.}(2016)\citenamefont
  {Chejanovsky}, \citenamefont {Rezai}, \citenamefont {Paolucci}, \citenamefont
  {Kim}, \citenamefont {Rendler}, \citenamefont {Rouabeh}, \citenamefont
  {F{\'a}varo~de Oliveira}, \citenamefont {Herlinger}, \citenamefont
  {Denisenko}, \citenamefont {Yang}, \citenamefont {Gerhardt}, \citenamefont
  {Finkler}, \citenamefont {Smet},\ and\ \citenamefont
  {Wrachtrup}}]{10.1021/acs.nanolett.6b03268}%
  \BibitemOpen
  \bibfield  {author} {\bibinfo {author} {\bibfnamefont {Nathan}\ \bibnamefont
  {Chejanovsky}}, \bibinfo {author} {\bibfnamefont {Mohammad}\ \bibnamefont
  {Rezai}}, \bibinfo {author} {\bibfnamefont {Federico}\ \bibnamefont
  {Paolucci}}, \bibinfo {author} {\bibfnamefont {Youngwook}\ \bibnamefont
  {Kim}}, \bibinfo {author} {\bibfnamefont {Torsten}\ \bibnamefont {Rendler}},
  \bibinfo {author} {\bibfnamefont {Wafa}\ \bibnamefont {Rouabeh}}, \bibinfo
  {author} {\bibfnamefont {Felipe}\ \bibnamefont {F{\'a}varo~de Oliveira}},
  \bibinfo {author} {\bibfnamefont {Patrick}\ \bibnamefont {Herlinger}},
  \bibinfo {author} {\bibfnamefont {Andrej}\ \bibnamefont {Denisenko}},
  \bibinfo {author} {\bibfnamefont {Sen}\ \bibnamefont {Yang}}, \bibinfo
  {author} {\bibfnamefont {Ilja}\ \bibnamefont {Gerhardt}}, \bibinfo {author}
  {\bibfnamefont {Amit}\ \bibnamefont {Finkler}}, \bibinfo {author}
  {\bibfnamefont {Jurgen~H.}\ \bibnamefont {Smet}}, \ and\ \bibinfo {author}
  {\bibfnamefont {Jörg}\ \bibnamefont {Wrachtrup}},\ }\bibfield  {title}
  {\enquote {\bibinfo {title} {Structural attributes and photodynamics of
  visible spectrum quantum emitters in hexagonal boron nitride},}\ }\href
  {\doibase 10.1021/acs.nanolett.6b03268} {\bibfield  {journal} {\bibinfo
  {journal} {Nano Lett.}\ }\textbf {\bibinfo {volume} {16}},\ \bibinfo {pages}
  {7037--7045} (\bibinfo {year} {2016})}\BibitemShut {NoStop}%
\bibitem [{\citenamefont {Xu}\ \emph {et~al.}(2018)\citenamefont {Xu},
  \citenamefont {Elbadawi}, \citenamefont {Tran}, \citenamefont {Kianinia},
  \citenamefont {Li}, \citenamefont {Liu}, \citenamefont {Hoffman},
  \citenamefont {Nguyen}, \citenamefont {Kim}, \citenamefont {Edgar},
  \citenamefont {Wu}, \citenamefont {Song}, \citenamefont {Ali}, \citenamefont
  {Ford}, \citenamefont {Toth},\ and\ \citenamefont
  {Aharonovich}}]{10.1039/C7NR08222C}%
  \BibitemOpen
  \bibfield  {author} {\bibinfo {author} {\bibfnamefont {Zai-Quan}\
  \bibnamefont {Xu}}, \bibinfo {author} {\bibfnamefont {Christopher}\
  \bibnamefont {Elbadawi}}, \bibinfo {author} {\bibfnamefont {Toan~Trong}\
  \bibnamefont {Tran}}, \bibinfo {author} {\bibfnamefont {Mehran}\ \bibnamefont
  {Kianinia}}, \bibinfo {author} {\bibfnamefont {Xiuling}\ \bibnamefont {Li}},
  \bibinfo {author} {\bibfnamefont {Daobin}\ \bibnamefont {Liu}}, \bibinfo
  {author} {\bibfnamefont {Timothy~B.}\ \bibnamefont {Hoffman}}, \bibinfo
  {author} {\bibfnamefont {Minh}\ \bibnamefont {Nguyen}}, \bibinfo {author}
  {\bibfnamefont {Sejeong}\ \bibnamefont {Kim}}, \bibinfo {author}
  {\bibfnamefont {James~H.}\ \bibnamefont {Edgar}}, \bibinfo {author}
  {\bibfnamefont {Xiaojun}\ \bibnamefont {Wu}}, \bibinfo {author}
  {\bibfnamefont {Li}~\bibnamefont {Song}}, \bibinfo {author} {\bibfnamefont
  {Sajid}\ \bibnamefont {Ali}}, \bibinfo {author} {\bibfnamefont {Mike}\
  \bibnamefont {Ford}}, \bibinfo {author} {\bibfnamefont {Milos}\ \bibnamefont
  {Toth}}, \ and\ \bibinfo {author} {\bibfnamefont {Igor}\ \bibnamefont
  {Aharonovich}},\ }\bibfield  {title} {\enquote {\bibinfo {title} {Single
  photon emission from plasma treated {2D} hexagonal boron nitride},}\ }\href
  {\doibase 10.1039/C7NR08222C} {\bibfield  {journal} {\bibinfo  {journal}
  {Nanoscale}\ }\textbf {\bibinfo {volume} {10}},\ \bibinfo {pages}
  {7957--7965} (\bibinfo {year} {2018})}\BibitemShut {NoStop}%
\bibitem [{\citenamefont {Choi}\ \emph {et~al.}(2016)\citenamefont {Choi},
  \citenamefont {Tran}, \citenamefont {Elbadawi}, \citenamefont {Lobo},
  \citenamefont {Wang}, \citenamefont {Juodkazis}, \citenamefont {Seniutinas},
  \citenamefont {Toth},\ and\ \citenamefont
  {Aharonovich}}]{10.1021/acsami.6b09875}%
  \BibitemOpen
  \bibfield  {author} {\bibinfo {author} {\bibfnamefont {Sumin}\ \bibnamefont
  {Choi}}, \bibinfo {author} {\bibfnamefont {Toan~Trong}\ \bibnamefont {Tran}},
  \bibinfo {author} {\bibfnamefont {Christopher}\ \bibnamefont {Elbadawi}},
  \bibinfo {author} {\bibfnamefont {Charlene}\ \bibnamefont {Lobo}}, \bibinfo
  {author} {\bibfnamefont {Xuewen}\ \bibnamefont {Wang}}, \bibinfo {author}
  {\bibfnamefont {Saulius}\ \bibnamefont {Juodkazis}}, \bibinfo {author}
  {\bibfnamefont {Gediminas}\ \bibnamefont {Seniutinas}}, \bibinfo {author}
  {\bibfnamefont {Milos}\ \bibnamefont {Toth}}, \ and\ \bibinfo {author}
  {\bibfnamefont {Igor}\ \bibnamefont {Aharonovich}},\ }\bibfield  {title}
  {\enquote {\bibinfo {title} {Engineering and localization of quantum emitters
  in large hexagonal boron nitride layers},}\ }\href {\doibase
  10.1021/acsami.6b09875} {\bibfield  {journal} {\bibinfo  {journal} {ACS Appl.
  Mater. Interfaces}\ }\textbf {\bibinfo {volume} {8}},\ \bibinfo {pages}
  {29642--29648} (\bibinfo {year} {2016})}\BibitemShut {NoStop}%
\bibitem [{\citenamefont {Ngoc My~Duong}\ \emph {et~al.}(2018)\citenamefont
  {Ngoc My~Duong}, \citenamefont {Nguyen}, \citenamefont {Kianinia},
  \citenamefont {Ohshima}, \citenamefont {Abe}, \citenamefont {Watanabe},
  \citenamefont {Taniguchi}, \citenamefont {Edgar}, \citenamefont
  {Aharonovich},\ and\ \citenamefont {Toth}}]{10.1021/acsami.8b07506}%
  \BibitemOpen
  \bibfield  {author} {\bibinfo {author} {\bibfnamefont {Hanh}\ \bibnamefont
  {Ngoc My~Duong}}, \bibinfo {author} {\bibfnamefont {Minh Anh~Phan}\
  \bibnamefont {Nguyen}}, \bibinfo {author} {\bibfnamefont {Mehran}\
  \bibnamefont {Kianinia}}, \bibinfo {author} {\bibfnamefont {Takeshi}\
  \bibnamefont {Ohshima}}, \bibinfo {author} {\bibfnamefont {Hiroshi}\
  \bibnamefont {Abe}}, \bibinfo {author} {\bibfnamefont {Kenji}\ \bibnamefont
  {Watanabe}}, \bibinfo {author} {\bibfnamefont {Takashi}\ \bibnamefont
  {Taniguchi}}, \bibinfo {author} {\bibfnamefont {James~H.}\ \bibnamefont
  {Edgar}}, \bibinfo {author} {\bibfnamefont {Igor}\ \bibnamefont
  {Aharonovich}}, \ and\ \bibinfo {author} {\bibfnamefont {Milos}\ \bibnamefont
  {Toth}},\ }\bibfield  {title} {\enquote {\bibinfo {title} {Effects of
  high-energy electron irradiation on quantum emitters in hexagonal boron
  nitride},}\ }\href {\doibase 10.1021/acsami.8b07506} {\bibfield  {journal}
  {\bibinfo  {journal} {ACS Appl. Mater. Interfaces}\ }\textbf {\bibinfo
  {volume} {10}},\ \bibinfo {pages} {24886--24891} (\bibinfo {year}
  {2018})}\BibitemShut {NoStop}%
\bibitem [{\citenamefont {Proscia}\ \emph {et~al.}(2018)\citenamefont
  {Proscia}, \citenamefont {Shotan}, \citenamefont {Jayakumar}, \citenamefont
  {Reddy}, \citenamefont {Cohen}, \citenamefont {Dollar}, \citenamefont
  {Alkauskas}, \citenamefont {Doherty}, \citenamefont {Meriles},\ and\
  \citenamefont {Menon}}]{10.1364/OPTICA.5.001128}%
  \BibitemOpen
  \bibfield  {author} {\bibinfo {author} {\bibfnamefont {Nicholas~V.}\
  \bibnamefont {Proscia}}, \bibinfo {author} {\bibfnamefont {Zav}\ \bibnamefont
  {Shotan}}, \bibinfo {author} {\bibfnamefont {Harishankar}\ \bibnamefont
  {Jayakumar}}, \bibinfo {author} {\bibfnamefont {Prithvi}\ \bibnamefont
  {Reddy}}, \bibinfo {author} {\bibfnamefont {Charles}\ \bibnamefont {Cohen}},
  \bibinfo {author} {\bibfnamefont {Michael}\ \bibnamefont {Dollar}}, \bibinfo
  {author} {\bibfnamefont {Audrius}\ \bibnamefont {Alkauskas}}, \bibinfo
  {author} {\bibfnamefont {Marcus}\ \bibnamefont {Doherty}}, \bibinfo {author}
  {\bibfnamefont {Carlos~A.}\ \bibnamefont {Meriles}}, \ and\ \bibinfo {author}
  {\bibfnamefont {Vinod~M.}\ \bibnamefont {Menon}},\ }\bibfield  {title}
  {\enquote {\bibinfo {title} {Near-deterministic activation of
  room-temperature quantum emitters in hexagonal boron nitride},}\ }\href
  {\doibase 10.1364/OPTICA.5.001128} {\bibfield  {journal} {\bibinfo  {journal}
  {Optica}\ }\textbf {\bibinfo {volume} {5}},\ \bibinfo {pages} {1128--1134}
  (\bibinfo {year} {2018})}\BibitemShut {NoStop}%
\bibitem [{\citenamefont {Gisin}\ \emph {et~al.}(2002)\citenamefont {Gisin},
  \citenamefont {Ribordy}, \citenamefont {Tittel},\ and\ \citenamefont
  {Zbinden}}]{10.1103/RevModPhys.74.145}%
  \BibitemOpen
  \bibfield  {author} {\bibinfo {author} {\bibfnamefont {Nicolas}\ \bibnamefont
  {Gisin}}, \bibinfo {author} {\bibfnamefont {Gr\'egoire}\ \bibnamefont
  {Ribordy}}, \bibinfo {author} {\bibfnamefont {Wolfgang}\ \bibnamefont
  {Tittel}}, \ and\ \bibinfo {author} {\bibfnamefont {Hugo}\ \bibnamefont
  {Zbinden}},\ }\bibfield  {title} {\enquote {\bibinfo {title} {Quantum
  cryptography},}\ }\href {\doibase 10.1103/RevModPhys.74.145} {\bibfield
  {journal} {\bibinfo  {journal} {Rev. Mod. Phys.}\ }\textbf {\bibinfo {volume}
  {74}},\ \bibinfo {pages} {145--195} (\bibinfo {year} {2002})}\BibitemShut
  {NoStop}%
\bibitem [{\citenamefont {Kok}\ \emph {et~al.}(2007)\citenamefont {Kok},
  \citenamefont {Munro}, \citenamefont {Nemoto}, \citenamefont {Ralph},
  \citenamefont {Dowling},\ and\ \citenamefont
  {Milburn}}]{10.1103/RevModPhys.79.135}%
  \BibitemOpen
  \bibfield  {author} {\bibinfo {author} {\bibfnamefont {Pieter}\ \bibnamefont
  {Kok}}, \bibinfo {author} {\bibfnamefont {W.~J.}\ \bibnamefont {Munro}},
  \bibinfo {author} {\bibfnamefont {Kae}\ \bibnamefont {Nemoto}}, \bibinfo
  {author} {\bibfnamefont {T.~C.}\ \bibnamefont {Ralph}}, \bibinfo {author}
  {\bibfnamefont {Jonathan~P.}\ \bibnamefont {Dowling}}, \ and\ \bibinfo
  {author} {\bibfnamefont {G.~J.}\ \bibnamefont {Milburn}},\ }\bibfield
  {title} {\enquote {\bibinfo {title} {Linear optical quantum computing with
  photonic qubits},}\ }\href {\doibase 10.1103/RevModPhys.79.135} {\bibfield
  {journal} {\bibinfo  {journal} {Rev. Mod. Phys.}\ }\textbf {\bibinfo {volume}
  {79}},\ \bibinfo {pages} {135--174} (\bibinfo {year} {2007})}\BibitemShut
  {NoStop}%
\bibitem [{\citenamefont {Vahala}(2003)}]{10.1038/nature01939}%
  \BibitemOpen
  \bibfield  {author} {\bibinfo {author} {\bibfnamefont {Kerry~J.}\
  \bibnamefont {Vahala}},\ }\bibfield  {title} {\enquote {\bibinfo {title}
  {Optical microcavities},}\ }\href {\doibase 10.1038/nature01939} {\bibfield
  {journal} {\bibinfo  {journal} {Nature}\ }\textbf {\bibinfo {volume} {424}},\
  \bibinfo {pages} {839--846} (\bibinfo {year} {2003})}\BibitemShut {NoStop}%
\bibitem [{\citenamefont {Kaupp}\ \emph {et~al.}(2013)\citenamefont {Kaupp},
  \citenamefont {Deutsch}, \citenamefont {Chang}, \citenamefont {Reichel},
  \citenamefont {H\"ansch},\ and\ \citenamefont
  {Hunger}}]{10.1103/PhysRevA.88.053812}%
  \BibitemOpen
  \bibfield  {author} {\bibinfo {author} {\bibfnamefont {Hanno}\ \bibnamefont
  {Kaupp}}, \bibinfo {author} {\bibfnamefont {Christian}\ \bibnamefont
  {Deutsch}}, \bibinfo {author} {\bibfnamefont {Huan-Cheng}\ \bibnamefont
  {Chang}}, \bibinfo {author} {\bibfnamefont {Jakob}\ \bibnamefont {Reichel}},
  \bibinfo {author} {\bibfnamefont {Theodor~W.}\ \bibnamefont {H\"ansch}}, \
  and\ \bibinfo {author} {\bibfnamefont {David}\ \bibnamefont {Hunger}},\
  }\bibfield  {title} {\enquote {\bibinfo {title} {Scaling laws of the cavity
  enhancement for nitrogen-vacancy centers in diamond},}\ }\href {\doibase
  10.1103/PhysRevA.88.053812} {\bibfield  {journal} {\bibinfo  {journal} {Phys.
  Rev. A}\ }\textbf {\bibinfo {volume} {88}},\ \bibinfo {pages} {053812}
  (\bibinfo {year} {2013})}\BibitemShut {NoStop}%
\bibitem [{\citenamefont {Iff}\ \emph {et~al.}(2018)\citenamefont {Iff},
  \citenamefont {Lundt}, \citenamefont {Betzold}, \citenamefont {Tripathi},
  \citenamefont {Emmerling}, \citenamefont {Tongay}, \citenamefont {Lee},
  \citenamefont {Kwon}, \citenamefont {H\"{o}fling},\ and\ \citenamefont
  {Schneider}}]{10.1364/OE.26.025944}%
  \BibitemOpen
  \bibfield  {author} {\bibinfo {author} {\bibfnamefont {Oliver}\ \bibnamefont
  {Iff}}, \bibinfo {author} {\bibfnamefont {Nils}\ \bibnamefont {Lundt}},
  \bibinfo {author} {\bibfnamefont {Simon}\ \bibnamefont {Betzold}}, \bibinfo
  {author} {\bibfnamefont {Laxmi~Narayan}\ \bibnamefont {Tripathi}}, \bibinfo
  {author} {\bibfnamefont {Monika}\ \bibnamefont {Emmerling}}, \bibinfo
  {author} {\bibfnamefont {Sefaattin}\ \bibnamefont {Tongay}}, \bibinfo
  {author} {\bibfnamefont {Young~Jin}\ \bibnamefont {Lee}}, \bibinfo {author}
  {\bibfnamefont {Soon-Hong}\ \bibnamefont {Kwon}}, \bibinfo {author}
  {\bibfnamefont {Sven}\ \bibnamefont {H\"{o}fling}}, \ and\ \bibinfo {author}
  {\bibfnamefont {Christian}\ \bibnamefont {Schneider}},\ }\bibfield  {title}
  {\enquote {\bibinfo {title} {Deterministic coupling of quantum emitters in
  {WSe2} monolayers to plasmonic nanocavities},}\ }\href {\doibase
  10.1364/OE.26.025944} {\bibfield  {journal} {\bibinfo  {journal} {Opt.
  Express}\ }\textbf {\bibinfo {volume} {26}},\ \bibinfo {pages} {25944--25951}
  (\bibinfo {year} {2018})}\BibitemShut {NoStop}%
\bibitem [{\citenamefont {Luo}\ \emph {et~al.}(2018)\citenamefont {Luo},
  \citenamefont {Shepard}, \citenamefont {Ardelean}, \citenamefont {Rhodes},
  \citenamefont {Kim}, \citenamefont {Barmak}, \citenamefont {Hone},\ and\
  \citenamefont {Strauf}}]{10.1038/s41565-018-0275-z}%
  \BibitemOpen
  \bibfield  {author} {\bibinfo {author} {\bibfnamefont {Yue}\ \bibnamefont
  {Luo}}, \bibinfo {author} {\bibfnamefont {Gabriella~D.}\ \bibnamefont
  {Shepard}}, \bibinfo {author} {\bibfnamefont {Jenny~V.}\ \bibnamefont
  {Ardelean}}, \bibinfo {author} {\bibfnamefont {Daniel~A.}\ \bibnamefont
  {Rhodes}}, \bibinfo {author} {\bibfnamefont {Bumho}\ \bibnamefont {Kim}},
  \bibinfo {author} {\bibfnamefont {Katayun}\ \bibnamefont {Barmak}}, \bibinfo
  {author} {\bibfnamefont {James~C.}\ \bibnamefont {Hone}}, \ and\ \bibinfo
  {author} {\bibfnamefont {Stefan}\ \bibnamefont {Strauf}},\ }\bibfield
  {title} {\enquote {\bibinfo {title} {Deterministic coupling of
  site-controlled quantum emitters in monolayer {WSe2} to plasmonic
  nanocavities},}\ }\href {\doibase 10.1038/s41565-018-0275-z} {\bibfield
  {journal} {\bibinfo  {journal} {Nat. Nanotechnol.}\ }\textbf {\bibinfo
  {volume} {13}},\ \bibinfo {pages} {1137--1142} (\bibinfo {year}
  {2018})}\BibitemShut {NoStop}%
\bibitem [{\citenamefont {Flatten}\ \emph {et~al.}(2018)\citenamefont
  {Flatten}, \citenamefont {Weng}, \citenamefont {Branny}, \citenamefont
  {Johnson}, \citenamefont {Dolan}, \citenamefont {Trichet}, \citenamefont
  {Gerardot},\ and\ \citenamefont {Smith}}]{10.1063/1.5026779}%
  \BibitemOpen
  \bibfield  {author} {\bibinfo {author} {\bibfnamefont {L.~C.}\ \bibnamefont
  {Flatten}}, \bibinfo {author} {\bibfnamefont {L.}~\bibnamefont {Weng}},
  \bibinfo {author} {\bibfnamefont {A.}~\bibnamefont {Branny}}, \bibinfo
  {author} {\bibfnamefont {S.}~\bibnamefont {Johnson}}, \bibinfo {author}
  {\bibfnamefont {P.~R.}\ \bibnamefont {Dolan}}, \bibinfo {author}
  {\bibfnamefont {A.~A.~P.}\ \bibnamefont {Trichet}}, \bibinfo {author}
  {\bibfnamefont {B.~D.}\ \bibnamefont {Gerardot}}, \ and\ \bibinfo {author}
  {\bibfnamefont {J.~M.}\ \bibnamefont {Smith}},\ }\bibfield  {title} {\enquote
  {\bibinfo {title} {Microcavity enhanced single photon emission from
  two-dimensional {WSe2}},}\ }\href {\doibase 10.1063/1.5026779} {\bibfield
  {journal} {\bibinfo  {journal} {Appl. Phys. Lett.}\ }\textbf {\bibinfo
  {volume} {112}},\ \bibinfo {pages} {191105} (\bibinfo {year}
  {2018})}\BibitemShut {NoStop}%
\bibitem [{\citenamefont {Tran}\ \emph {et~al.}(2017)\citenamefont {Tran},
  \citenamefont {Wang}, \citenamefont {Xu}, \citenamefont {Yang}, \citenamefont
  {Toth}, \citenamefont {Odom},\ and\ \citenamefont
  {Aharonovich}}]{doi:10.1021/acs.nanolett.7b00444}%
  \BibitemOpen
  \bibfield  {author} {\bibinfo {author} {\bibfnamefont {Toan~Trong}\
  \bibnamefont {Tran}}, \bibinfo {author} {\bibfnamefont {Danqing}\
  \bibnamefont {Wang}}, \bibinfo {author} {\bibfnamefont {Zai-Quan}\
  \bibnamefont {Xu}}, \bibinfo {author} {\bibfnamefont {Ankun}\ \bibnamefont
  {Yang}}, \bibinfo {author} {\bibfnamefont {Milos}\ \bibnamefont {Toth}},
  \bibinfo {author} {\bibfnamefont {Teri~W.}\ \bibnamefont {Odom}}, \ and\
  \bibinfo {author} {\bibfnamefont {Igor}\ \bibnamefont {Aharonovich}},\
  }\bibfield  {title} {\enquote {\bibinfo {title} {Deterministic coupling of
  quantum emitters in 2d materials to plasmonic nanocavity arrays},}\ }\href
  {\doibase 10.1021/acs.nanolett.7b00444} {\bibfield  {journal} {\bibinfo
  {journal} {Nano Lett.}\ }\textbf {\bibinfo {volume} {17}},\ \bibinfo {pages}
  {2634--2639} (\bibinfo {year} {2017})}\BibitemShut {NoStop}%
\bibitem [{\citenamefont {Kim}\ \emph {et~al.}(2018)\citenamefont {Kim},
  \citenamefont {Fr{\"o}ch}, \citenamefont {Christian}, \citenamefont {Straw},
  \citenamefont {Bishop}, \citenamefont {Totonjian}, \citenamefont {Watanabe},
  \citenamefont {Taniguchi}, \citenamefont {Toth},\ and\ \citenamefont
  {Aharonovich}}]{10.1038/s41467-018-05117-4}%
  \BibitemOpen
  \bibfield  {author} {\bibinfo {author} {\bibfnamefont {Sejeong}\ \bibnamefont
  {Kim}}, \bibinfo {author} {\bibfnamefont {Johannes~E.}\ \bibnamefont
  {Fr{\"o}ch}}, \bibinfo {author} {\bibfnamefont {Joe}\ \bibnamefont
  {Christian}}, \bibinfo {author} {\bibfnamefont {Marcus}\ \bibnamefont
  {Straw}}, \bibinfo {author} {\bibfnamefont {James}\ \bibnamefont {Bishop}},
  \bibinfo {author} {\bibfnamefont {Daniel}\ \bibnamefont {Totonjian}},
  \bibinfo {author} {\bibfnamefont {Kenji}\ \bibnamefont {Watanabe}}, \bibinfo
  {author} {\bibfnamefont {Takashi}\ \bibnamefont {Taniguchi}}, \bibinfo
  {author} {\bibfnamefont {Milos}\ \bibnamefont {Toth}}, \ and\ \bibinfo
  {author} {\bibfnamefont {Igor}\ \bibnamefont {Aharonovich}},\ }\bibfield
  {title} {\enquote {\bibinfo {title} {Photonic crystal cavities from hexagonal
  boron nitride},}\ }\href {\doibase 10.1038/s41467-018-05117-4} {\bibfield
  {journal} {\bibinfo  {journal} {Nat. Commun.}\ }\textbf {\bibinfo {volume}
  {9}},\ \bibinfo {pages} {2623} (\bibinfo {year} {2018})}\BibitemShut
  {NoStop}%
\bibitem [{\citenamefont {Liao}\ \emph {et~al.}(2017)\citenamefont {Liao},
  \citenamefont {Cai}, \citenamefont {Liu}, \citenamefont {Zhang},
  \citenamefont {Li}, \citenamefont {Ren}, \citenamefont {Yin}, \citenamefont
  {Shen}, \citenamefont {Cao}, \citenamefont {Li}, \citenamefont {Li},
  \citenamefont {Chen}, \citenamefont {Sun}, \citenamefont {Jia}, \citenamefont
  {Wu}, \citenamefont {Jiang}, \citenamefont {Wang}, \citenamefont {Huang},
  \citenamefont {Wang}, \citenamefont {Zhou}, \citenamefont {Deng},
  \citenamefont {Xi}, \citenamefont {Ma}, \citenamefont {Hu}, \citenamefont
  {Zhang}, \citenamefont {Chen}, \citenamefont {Liu}, \citenamefont {Wang},
  \citenamefont {Zhu}, \citenamefont {Lu}, \citenamefont {Shu}, \citenamefont
  {Peng}, \citenamefont {Wang},\ and\ \citenamefont
  {Pan}}]{10.1038/nature23655}%
  \BibitemOpen
  \bibfield  {author} {\bibinfo {author} {\bibfnamefont {Sheng-Kai}\
  \bibnamefont {Liao}}, \bibinfo {author} {\bibfnamefont {Wen-Qi}\ \bibnamefont
  {Cai}}, \bibinfo {author} {\bibfnamefont {Wei-Yue}\ \bibnamefont {Liu}},
  \bibinfo {author} {\bibfnamefont {Liang}\ \bibnamefont {Zhang}}, \bibinfo
  {author} {\bibfnamefont {Yang}\ \bibnamefont {Li}}, \bibinfo {author}
  {\bibfnamefont {Ji-Gang}\ \bibnamefont {Ren}}, \bibinfo {author}
  {\bibfnamefont {Juan}\ \bibnamefont {Yin}}, \bibinfo {author} {\bibfnamefont
  {Qi}~\bibnamefont {Shen}}, \bibinfo {author} {\bibfnamefont {Yuan}\
  \bibnamefont {Cao}}, \bibinfo {author} {\bibfnamefont {Zheng-Ping}\
  \bibnamefont {Li}}, \bibinfo {author} {\bibfnamefont {Feng-Zhi}\ \bibnamefont
  {Li}}, \bibinfo {author} {\bibfnamefont {Xia-Wei}\ \bibnamefont {Chen}},
  \bibinfo {author} {\bibfnamefont {Li-Hua}\ \bibnamefont {Sun}}, \bibinfo
  {author} {\bibfnamefont {Jian-Jun}\ \bibnamefont {Jia}}, \bibinfo {author}
  {\bibfnamefont {Jin-Cai}\ \bibnamefont {Wu}}, \bibinfo {author}
  {\bibfnamefont {Xiao-Jun}\ \bibnamefont {Jiang}}, \bibinfo {author}
  {\bibfnamefont {Jian-Feng}\ \bibnamefont {Wang}}, \bibinfo {author}
  {\bibfnamefont {Yong-Mei}\ \bibnamefont {Huang}}, \bibinfo {author}
  {\bibfnamefont {Qiang}\ \bibnamefont {Wang}}, \bibinfo {author}
  {\bibfnamefont {Yi-Lin}\ \bibnamefont {Zhou}}, \bibinfo {author}
  {\bibfnamefont {Lei}\ \bibnamefont {Deng}}, \bibinfo {author} {\bibfnamefont
  {Tao}\ \bibnamefont {Xi}}, \bibinfo {author} {\bibfnamefont {Lu}~\bibnamefont
  {Ma}}, \bibinfo {author} {\bibfnamefont {Tai}\ \bibnamefont {Hu}}, \bibinfo
  {author} {\bibfnamefont {Qiang}\ \bibnamefont {Zhang}}, \bibinfo {author}
  {\bibfnamefont {Yu-Ao}\ \bibnamefont {Chen}}, \bibinfo {author}
  {\bibfnamefont {Nai-Le}\ \bibnamefont {Liu}}, \bibinfo {author}
  {\bibfnamefont {Xiang-Bin}\ \bibnamefont {Wang}}, \bibinfo {author}
  {\bibfnamefont {Zhen-Cai}\ \bibnamefont {Zhu}}, \bibinfo {author}
  {\bibfnamefont {Chao-Yang}\ \bibnamefont {Lu}}, \bibinfo {author}
  {\bibfnamefont {Rong}\ \bibnamefont {Shu}}, \bibinfo {author} {\bibfnamefont
  {Cheng-Zhi}\ \bibnamefont {Peng}}, \bibinfo {author} {\bibfnamefont
  {Jian-Yu}\ \bibnamefont {Wang}}, \ and\ \bibinfo {author} {\bibfnamefont
  {Jian-Wei}\ \bibnamefont {Pan}},\ }\bibfield  {title} {\enquote {\bibinfo
  {title} {Satellite-to-ground quantum key distribution},}\ }\href {\doibase
  10.1038/nature23655} {\bibfield  {journal} {\bibinfo  {journal} {Nature}\
  }\textbf {\bibinfo {volume} {549}},\ \bibinfo {pages} {43--47} (\bibinfo
  {year} {2017})}\BibitemShut {NoStop}%
\bibitem [{\citenamefont {Dolan}\ \emph {et~al.}(2010)\citenamefont {Dolan},
  \citenamefont {Hughes}, \citenamefont {Grazioso}, \citenamefont {Patton},\
  and\ \citenamefont {Smith}}]{10.1364/OL.35.003556}%
  \BibitemOpen
  \bibfield  {author} {\bibinfo {author} {\bibfnamefont {Philip~R.}\
  \bibnamefont {Dolan}}, \bibinfo {author} {\bibfnamefont {Gareth~M.}\
  \bibnamefont {Hughes}}, \bibinfo {author} {\bibfnamefont {Fabio}\
  \bibnamefont {Grazioso}}, \bibinfo {author} {\bibfnamefont {Brian~R.}\
  \bibnamefont {Patton}}, \ and\ \bibinfo {author} {\bibfnamefont {Jason~M.}\
  \bibnamefont {Smith}},\ }\bibfield  {title} {\enquote {\bibinfo {title}
  {Femtoliter tunable optical cavity arrays},}\ }\href {\doibase
  10.1364/OL.35.003556} {\bibfield  {journal} {\bibinfo  {journal} {Opt.
  Lett.}\ }\textbf {\bibinfo {volume} {35}},\ \bibinfo {pages} {3556--3558}
  (\bibinfo {year} {2010})}\BibitemShut {NoStop}%
\bibitem [{\citenamefont {Trichet}\ \emph {et~al.}(2015)\citenamefont
  {Trichet}, \citenamefont {Dolan}, \citenamefont {Coles}, \citenamefont
  {Hughes},\ and\ \citenamefont {Smith}}]{10.1364/OE.23.017205}%
  \BibitemOpen
  \bibfield  {author} {\bibinfo {author} {\bibfnamefont {Aur\'{e}lien A.~P.}\
  \bibnamefont {Trichet}}, \bibinfo {author} {\bibfnamefont {Philip~R.}\
  \bibnamefont {Dolan}}, \bibinfo {author} {\bibfnamefont {David~M.}\
  \bibnamefont {Coles}}, \bibinfo {author} {\bibfnamefont {Gareth~M.}\
  \bibnamefont {Hughes}}, \ and\ \bibinfo {author} {\bibfnamefont {Jason~M.}\
  \bibnamefont {Smith}},\ }\bibfield  {title} {\enquote {\bibinfo {title}
  {Topographic control of open-access microcavities at the nanometer scale},}\
  }\href {\doibase 10.1364/OE.23.017205} {\bibfield  {journal} {\bibinfo
  {journal} {Opt. Express}\ }\textbf {\bibinfo {volume} {23}},\ \bibinfo
  {pages} {17205--17216} (\bibinfo {year} {2015})}\BibitemShut {NoStop}%
\bibitem [{sup()}]{supp_mat}%
  \BibitemOpen
  \href@noop {} {}\bibinfo {note} {See Supplemental Material at [URL will be
  inserted by publisher] for details on hemisphere fabrication, spin speed
  curves of the PDMS solution, photographs of the device, transverse mode
  spacing, FDTD simulations and QKD simulations}\BibitemShut {NoStop}%
\bibitem [{\citenamefont {Liu}\ \emph {et~al.}(2014)\citenamefont {Liu},
  \citenamefont {Galfsky}, \citenamefont {Sun}, \citenamefont {Xia},
  \citenamefont {Lin}, \citenamefont {Lee}, \citenamefont {K{\'e}na-Cohen},\
  and\ \citenamefont {Menon}}]{10.1038/nphoton.2014.304}%
  \BibitemOpen
  \bibfield  {author} {\bibinfo {author} {\bibfnamefont {Xiaoze}\ \bibnamefont
  {Liu}}, \bibinfo {author} {\bibfnamefont {Tal}\ \bibnamefont {Galfsky}},
  \bibinfo {author} {\bibfnamefont {Zheng}\ \bibnamefont {Sun}}, \bibinfo
  {author} {\bibfnamefont {Fengnian}\ \bibnamefont {Xia}}, \bibinfo {author}
  {\bibfnamefont {Erh-chen}\ \bibnamefont {Lin}}, \bibinfo {author}
  {\bibfnamefont {Yi-Hsien}\ \bibnamefont {Lee}}, \bibinfo {author}
  {\bibfnamefont {St{\'e}phane}\ \bibnamefont {K{\'e}na-Cohen}}, \ and\
  \bibinfo {author} {\bibfnamefont {Vinod~M.}\ \bibnamefont {Menon}},\
  }\bibfield  {title} {\enquote {\bibinfo {title} {Strong light-matter coupling
  in two-dimensional atomic crystals},}\ }\href {\doibase
  10.1038/nphoton.2014.304} {\bibfield  {journal} {\bibinfo  {journal} {Nat.
  Photon.}\ }\textbf {\bibinfo {volume} {9}},\ \bibinfo {pages} {30--34}
  (\bibinfo {year} {2014})}\BibitemShut {NoStop}%
\bibitem [{\citenamefont {Lundt}\ \emph {et~al.}(2016)\citenamefont {Lundt},
  \citenamefont {Klembt}, \citenamefont {Cherotchenko}, \citenamefont
  {Betzold}, \citenamefont {Iff}, \citenamefont {Nalitov}, \citenamefont
  {Klaas}, \citenamefont {Dietrich}, \citenamefont {Kavokin}, \citenamefont
  {H{\"o}fling},\ and\ \citenamefont {Schneider}}]{10.1038/ncomms13328}%
  \BibitemOpen
  \bibfield  {author} {\bibinfo {author} {\bibfnamefont {Nils}\ \bibnamefont
  {Lundt}}, \bibinfo {author} {\bibfnamefont {Sebastian}\ \bibnamefont
  {Klembt}}, \bibinfo {author} {\bibfnamefont {Evgeniia}\ \bibnamefont
  {Cherotchenko}}, \bibinfo {author} {\bibfnamefont {Simon}\ \bibnamefont
  {Betzold}}, \bibinfo {author} {\bibfnamefont {Oliver}\ \bibnamefont {Iff}},
  \bibinfo {author} {\bibfnamefont {Anton~V.}\ \bibnamefont {Nalitov}},
  \bibinfo {author} {\bibfnamefont {Martin}\ \bibnamefont {Klaas}}, \bibinfo
  {author} {\bibfnamefont {Christof~P.}\ \bibnamefont {Dietrich}}, \bibinfo
  {author} {\bibfnamefont {Alexey~V.}\ \bibnamefont {Kavokin}}, \bibinfo
  {author} {\bibfnamefont {Sven}\ \bibnamefont {H{\"o}fling}}, \ and\ \bibinfo
  {author} {\bibfnamefont {Christian}\ \bibnamefont {Schneider}},\ }\bibfield
  {title} {\enquote {\bibinfo {title} {Room-temperature tamm-plasmon
  exciton-polaritons with a {WSe2} monolayer},}\ }\href {\doibase
  10.1038/ncomms13328} {\bibfield  {journal} {\bibinfo  {journal} {Nat.
  Commun.}\ }\textbf {\bibinfo {volume} {7}},\ \bibinfo {pages} {13328}
  (\bibinfo {year} {2016})}\BibitemShut {NoStop}%
\bibitem [{\citenamefont {Knopf}\ \emph {et~al.}(2019)\citenamefont {Knopf},
  \citenamefont {Lundt}, \citenamefont {Bucher}, \citenamefont {H\"{o}fling},
  \citenamefont {Tongay}, \citenamefont {Taniguchi}, \citenamefont {Watanabe},
  \citenamefont {Staude}, \citenamefont {Schulz}, \citenamefont {Schneider},\
  and\ \citenamefont {Eilenberger}}]{10.1364/OME.9.000598}%
  \BibitemOpen
  \bibfield  {author} {\bibinfo {author} {\bibfnamefont {Heiko}\ \bibnamefont
  {Knopf}}, \bibinfo {author} {\bibfnamefont {Nils}\ \bibnamefont {Lundt}},
  \bibinfo {author} {\bibfnamefont {Tobias}\ \bibnamefont {Bucher}}, \bibinfo
  {author} {\bibfnamefont {Sven}\ \bibnamefont {H\"{o}fling}}, \bibinfo
  {author} {\bibfnamefont {Sefaattin}\ \bibnamefont {Tongay}}, \bibinfo
  {author} {\bibfnamefont {Takashi}\ \bibnamefont {Taniguchi}}, \bibinfo
  {author} {\bibfnamefont {Kenji}\ \bibnamefont {Watanabe}}, \bibinfo {author}
  {\bibfnamefont {Isabelle}\ \bibnamefont {Staude}}, \bibinfo {author}
  {\bibfnamefont {Ulrike}\ \bibnamefont {Schulz}}, \bibinfo {author}
  {\bibfnamefont {Christian}\ \bibnamefont {Schneider}}, \ and\ \bibinfo
  {author} {\bibfnamefont {Falk}\ \bibnamefont {Eilenberger}},\ }\bibfield
  {title} {\enquote {\bibinfo {title} {{Integration of atomically thin layers
  of transition metal dichalcogenides into high-Q, monolithic Bragg-cavities:
  an experimental platform for the enhancement of the optical interaction in
  2D-materials}},}\ }\href {\doibase 10.1364/OME.9.000598} {\bibfield
  {journal} {\bibinfo  {journal} {Opt. Mater. Express}\ }\textbf {\bibinfo
  {volume} {9}},\ \bibinfo {pages} {598--610} (\bibinfo {year}
  {2019})}\BibitemShut {NoStop}%
\bibitem [{\citenamefont {Hanschke}\ \emph {et~al.}(2018)\citenamefont
  {Hanschke}, \citenamefont {Fischer}, \citenamefont {Appel}, \citenamefont
  {Lukin}, \citenamefont {Wierzbowski}, \citenamefont {Sun}, \citenamefont
  {Trivedi}, \citenamefont {Vu{\v c}kovi{\'c}}, \citenamefont {Finley},\ and\
  \citenamefont {M{\"u}ller}}]{10.1038/s41534-018-0092-0}%
  \BibitemOpen
  \bibfield  {author} {\bibinfo {author} {\bibfnamefont {Lukas}\ \bibnamefont
  {Hanschke}}, \bibinfo {author} {\bibfnamefont {Kevin~A.}\ \bibnamefont
  {Fischer}}, \bibinfo {author} {\bibfnamefont {Stefan}\ \bibnamefont {Appel}},
  \bibinfo {author} {\bibfnamefont {Daniil}\ \bibnamefont {Lukin}}, \bibinfo
  {author} {\bibfnamefont {Jakob}\ \bibnamefont {Wierzbowski}}, \bibinfo
  {author} {\bibfnamefont {Shuo}\ \bibnamefont {Sun}}, \bibinfo {author}
  {\bibfnamefont {Rahul}\ \bibnamefont {Trivedi}}, \bibinfo {author}
  {\bibfnamefont {Jelena}\ \bibnamefont {Vu{\v c}kovi{\'c}}}, \bibinfo {author}
  {\bibfnamefont {Jonathan~J.}\ \bibnamefont {Finley}}, \ and\ \bibinfo
  {author} {\bibfnamefont {Kai}\ \bibnamefont {M{\"u}ller}},\ }\bibfield
  {title} {\enquote {\bibinfo {title} {Quantum dot single-photon sources with
  ultra-low multi-photon probability},}\ }\href {\doibase
  10.1038/s41534-018-0092-0} {\bibfield  {journal} {\bibinfo  {journal} {npj
  Quantum Inf.}\ }\textbf {\bibinfo {volume} {4}},\ \bibinfo {pages} {43}
  (\bibinfo {year} {2018})}\BibitemShut {NoStop}%
\bibitem [{\citenamefont {Schweickert}\ \emph {et~al.}(2018)\citenamefont
  {Schweickert}, \citenamefont {J{\"o}ns}, \citenamefont {Zeuner},
  \citenamefont {Covre~da Silva}, \citenamefont {Huang}, \citenamefont
  {Lettner}, \citenamefont {Reindl}, \citenamefont {Zichi}, \citenamefont
  {Trotta}, \citenamefont {Rastelli},\ and\ \citenamefont
  {Zwiller}}]{10.1063/1.5020038}%
  \BibitemOpen
  \bibfield  {author} {\bibinfo {author} {\bibfnamefont {Lucas}\ \bibnamefont
  {Schweickert}}, \bibinfo {author} {\bibfnamefont {Klaus~D.}\ \bibnamefont
  {J{\"o}ns}}, \bibinfo {author} {\bibfnamefont {Katharina~D.}\ \bibnamefont
  {Zeuner}}, \bibinfo {author} {\bibfnamefont {Saimon~Filipe}\ \bibnamefont
  {Covre~da Silva}}, \bibinfo {author} {\bibfnamefont {Huiying}\ \bibnamefont
  {Huang}}, \bibinfo {author} {\bibfnamefont {Thomas}\ \bibnamefont {Lettner}},
  \bibinfo {author} {\bibfnamefont {Marcus}\ \bibnamefont {Reindl}}, \bibinfo
  {author} {\bibfnamefont {Julien}\ \bibnamefont {Zichi}}, \bibinfo {author}
  {\bibfnamefont {Rinaldo}\ \bibnamefont {Trotta}}, \bibinfo {author}
  {\bibfnamefont {Armando}\ \bibnamefont {Rastelli}}, \ and\ \bibinfo {author}
  {\bibfnamefont {Val}\ \bibnamefont {Zwiller}},\ }\bibfield  {title} {\enquote
  {\bibinfo {title} {On-demand generation of background-free single photons
  from a solid-state source},}\ }\href {\doibase 10.1063/1.5020038} {\bibfield
  {journal} {\bibinfo  {journal} {Appl. Phys. Lett.}\ }\textbf {\bibinfo
  {volume} {112}},\ \bibinfo {pages} {093106} (\bibinfo {year}
  {2018})}\BibitemShut {NoStop}%
\bibitem [{\citenamefont {Beveratos}\ \emph {et~al.}(2001)\citenamefont
  {Beveratos}, \citenamefont {Brouri}, \citenamefont {Gacoin}, \citenamefont
  {Poizat},\ and\ \citenamefont {Grangier}}]{PhysRevA.64.061802}%
  \BibitemOpen
  \bibfield  {author} {\bibinfo {author} {\bibfnamefont {Alexios}\ \bibnamefont
  {Beveratos}}, \bibinfo {author} {\bibfnamefont {Rosa}\ \bibnamefont
  {Brouri}}, \bibinfo {author} {\bibfnamefont {Thierry}\ \bibnamefont
  {Gacoin}}, \bibinfo {author} {\bibfnamefont {Jean-Philippe}\ \bibnamefont
  {Poizat}}, \ and\ \bibinfo {author} {\bibfnamefont {Philippe}\ \bibnamefont
  {Grangier}},\ }\bibfield  {title} {\enquote {\bibinfo {title} {Nonclassical
  radiation from diamond nanocrystals},}\ }\href {\doibase
  10.1103/PhysRevA.64.061802} {\bibfield  {journal} {\bibinfo  {journal} {Phys.
  Rev. A}\ }\textbf {\bibinfo {volume} {64}},\ \bibinfo {pages} {061802}
  (\bibinfo {year} {2001})}\BibitemShut {NoStop}%
\bibitem [{\citenamefont {Gobby}\ \emph {et~al.}(2004)\citenamefont {Gobby},
  \citenamefont {Yuan},\ and\ \citenamefont {Shields}}]{10.1063/1.1738173}%
  \BibitemOpen
  \bibfield  {author} {\bibinfo {author} {\bibfnamefont {C.}~\bibnamefont
  {Gobby}}, \bibinfo {author} {\bibfnamefont {Z.~L.}\ \bibnamefont {Yuan}}, \
  and\ \bibinfo {author} {\bibfnamefont {A.~J.}\ \bibnamefont {Shields}},\
  }\bibfield  {title} {\enquote {\bibinfo {title} {Quantum key distribution
  over 122 km of standard telecom fiber},}\ }\href {\doibase 10.1063/1.1738173}
  {\bibfield  {journal} {\bibinfo  {journal} {Appl. Phys. Lett.}\ }\textbf
  {\bibinfo {volume} {84}},\ \bibinfo {pages} {3762--3764} (\bibinfo {year}
  {2004})}\BibitemShut {NoStop}%
\bibitem [{\citenamefont {Lo}\ \emph {et~al.}(2005)\citenamefont {Lo},
  \citenamefont {Ma},\ and\ \citenamefont {Chen}}]{PhysRevLett.94.230504}%
  \BibitemOpen
  \bibfield  {author} {\bibinfo {author} {\bibfnamefont {Hoi-Kwong}\
  \bibnamefont {Lo}}, \bibinfo {author} {\bibfnamefont {Xiongfeng}\
  \bibnamefont {Ma}}, \ and\ \bibinfo {author} {\bibfnamefont {Kai}\
  \bibnamefont {Chen}},\ }\bibfield  {title} {\enquote {\bibinfo {title} {Decoy
  state quantum key distribution},}\ }\href {\doibase
  10.1103/PhysRevLett.94.230504} {\bibfield  {journal} {\bibinfo  {journal}
  {Phys. Rev. Lett.}\ }\textbf {\bibinfo {volume} {94}},\ \bibinfo {pages}
  {230504} (\bibinfo {year} {2005})}\BibitemShut {NoStop}%
\bibitem [{\citenamefont {Raussendorf}\ and\ \citenamefont
  {Briegel}(2001)}]{PhysRevLett.86.5188}%
  \BibitemOpen
  \bibfield  {author} {\bibinfo {author} {\bibfnamefont {Robert}\ \bibnamefont
  {Raussendorf}}\ and\ \bibinfo {author} {\bibfnamefont {Hans~J.}\ \bibnamefont
  {Briegel}},\ }\bibfield  {title} {\enquote {\bibinfo {title} {A one-way
  quantum computer},}\ }\href {\doibase 10.1103/PhysRevLett.86.5188} {\bibfield
   {journal} {\bibinfo  {journal} {Phys. Rev. Lett.}\ }\textbf {\bibinfo
  {volume} {86}},\ \bibinfo {pages} {5188--5191} (\bibinfo {year}
  {2001})}\BibitemShut {NoStop}%
\bibitem [{\citenamefont {Hong}\ \emph {et~al.}(1987)\citenamefont {Hong},
  \citenamefont {Ou},\ and\ \citenamefont {Mandel}}]{PhysRevLett.59.2044}%
  \BibitemOpen
  \bibfield  {author} {\bibinfo {author} {\bibfnamefont {C.~K.}\ \bibnamefont
  {Hong}}, \bibinfo {author} {\bibfnamefont {Z.~Y.}\ \bibnamefont {Ou}}, \ and\
  \bibinfo {author} {\bibfnamefont {L.}~\bibnamefont {Mandel}},\ }\bibfield
  {title} {\enquote {\bibinfo {title} {Measurement of subpicosecond time
  intervals between two photons by interference},}\ }\href {\doibase
  10.1103/PhysRevLett.59.2044} {\bibfield  {journal} {\bibinfo  {journal}
  {Phys. Rev. Lett.}\ }\textbf {\bibinfo {volume} {59}},\ \bibinfo {pages}
  {2044--2046} (\bibinfo {year} {1987})}\BibitemShut {NoStop}%
\bibitem [{\citenamefont {Grange}\ \emph {et~al.}(2015)\citenamefont {Grange},
  \citenamefont {Hornecker}, \citenamefont {Hunger}, \citenamefont {Poizat},
  \citenamefont {G\'erard}, \citenamefont {Senellart},\ and\ \citenamefont
  {Auff\`eves}}]{PhysRevLett.114.193601}%
  \BibitemOpen
  \bibfield  {author} {\bibinfo {author} {\bibfnamefont {Thomas}\ \bibnamefont
  {Grange}}, \bibinfo {author} {\bibfnamefont {Gaston}\ \bibnamefont
  {Hornecker}}, \bibinfo {author} {\bibfnamefont {David}\ \bibnamefont
  {Hunger}}, \bibinfo {author} {\bibfnamefont {Jean-Philippe}\ \bibnamefont
  {Poizat}}, \bibinfo {author} {\bibfnamefont {Jean-Michel}\ \bibnamefont
  {G\'erard}}, \bibinfo {author} {\bibfnamefont {Pascale}\ \bibnamefont
  {Senellart}}, \ and\ \bibinfo {author} {\bibfnamefont {Alexia}\ \bibnamefont
  {Auff\`eves}},\ }\bibfield  {title} {\enquote {\bibinfo {title}
  {Cavity-funneled generation of indistinguishable single photons from strongly
  dissipative quantum emitters},}\ }\href {\doibase
  10.1103/PhysRevLett.114.193601} {\bibfield  {journal} {\bibinfo  {journal}
  {Phys. Rev. Lett.}\ }\textbf {\bibinfo {volume} {114}},\ \bibinfo {pages}
  {193601} (\bibinfo {year} {2015})}\BibitemShut {NoStop}%
\bibitem [{\citenamefont {Dietrich}\ \emph {et~al.}(2018)\citenamefont
  {Dietrich}, \citenamefont {B\"urk}, \citenamefont {Steiger}, \citenamefont
  {Antoniuk}, \citenamefont {Tran}, \citenamefont {Nguyen}, \citenamefont
  {Aharonovich}, \citenamefont {Jelezko},\ and\ \citenamefont
  {Kubanek}}]{PhysRevB.98.081414}%
  \BibitemOpen
  \bibfield  {author} {\bibinfo {author} {\bibfnamefont {A.}~\bibnamefont
  {Dietrich}}, \bibinfo {author} {\bibfnamefont {M.}~\bibnamefont {B\"urk}},
  \bibinfo {author} {\bibfnamefont {E.~S.}\ \bibnamefont {Steiger}}, \bibinfo
  {author} {\bibfnamefont {L.}~\bibnamefont {Antoniuk}}, \bibinfo {author}
  {\bibfnamefont {T.~T.}\ \bibnamefont {Tran}}, \bibinfo {author}
  {\bibfnamefont {M.}~\bibnamefont {Nguyen}}, \bibinfo {author} {\bibfnamefont
  {I.}~\bibnamefont {Aharonovich}}, \bibinfo {author} {\bibfnamefont
  {F.}~\bibnamefont {Jelezko}}, \ and\ \bibinfo {author} {\bibfnamefont
  {A.}~\bibnamefont {Kubanek}},\ }\bibfield  {title} {\enquote {\bibinfo
  {title} {Observation of fourier transform limited lines in hexagonal boron
  nitride},}\ }\href {\doibase 10.1103/PhysRevB.98.081414} {\bibfield
  {journal} {\bibinfo  {journal} {Phys. Rev. B}\ }\textbf {\bibinfo {volume}
  {98}},\ \bibinfo {pages} {081414} (\bibinfo {year} {2018})}\BibitemShut
  {NoStop}%
\end{thebibliography}
%

\clearpage

\end{document}